\documentclass[aps,journal=prb,twocolumn,noeprint]{revtex4-1}
\usepackage{graphicx}
\usepackage{epstopdf}
\usepackage{bmpsize}
\usepackage{amsmath,amsfonts,amsthm,bm}
\usepackage{bigints}
\usepackage{xcolor}
\usepackage{tikz}
\usetikzlibrary{decorations.pathreplacing}
\usepackage{ragged2e}

\usepackage[caption=false]{subfig}
\newcommand{\phantomsubfloat}[1]{
    {
        \captionsetup[subfigure]{labelformat=empty}
        \subfloat[][]{#1}
    }%
}

\usepackage{comment} 
\usepackage[bookmarks=false]{hyperref}
\hypersetup{colorlinks=true, citecolor=blue, urlcolor=blue, linkcolor=blue}

\usepackage{mathtools}

\newcommand\vertarrowbox[3][6ex]{%
  \begin{array}[t]{@{}c@{}} #2 \\
  \left\uparrow\vcenter{\hrule height #1}\right.\kern-\nulldelimiterspace\\
  \makebox[0pt]{\scriptsize#3}
  \end{array}%
}

\makeatletter
\renewcommand{\maketag@@@}[1]{\hbox{\m@th\normalsize\normalfont#1}}%
\makeatother

\newcommand{\ph}{{\phantom a}}
\def\!{\mskip-\thinmuskip}

\begin{document}
\title{Nonadiabatic Dynamics of Molecules Interacting with Metal Surfaces: Extending the Hierarchical Equations of Motion and Langevin Dynamics Approach to Position-Dependent Metal-Molecule Couplings}
\author{Martin Mäck}
\email[Corresponding author: Martin Mäck \\ Email:
]{martin.maeck@physik.uni-freiburg.de}
\affiliation{Institute of Physics, University of Freiburg, Hermann-Herder-Strasse 3, 79104 Freiburg, Germany}

\author{Samuel L. Rudge}
\affiliation{Institute of Physics, University of Freiburg, Hermann-Herder-Strasse 3, 79104 Freiburg, Germany}
\author{Michael Thoss}
\affiliation{Institute of Physics, University of Freiburg, Hermann-Herder-Strasse 3, 79104 Freiburg, Germany}

\begin{abstract}
\noindent 
Electronic friction and Langevin dynamics is a popular mixed quantum-classical method for simulating the nonadiabatic dynamics of molecules interacting with metal surfaces, as it can be computationally more efficient than fully quantum approaches. Previous approaches to calculating the electronic friction and other forces, however, have been limited to either noninteracting molecular models or position-independent metal-molecule couplings. In this work, we extend the theory of electronic friction within the hierarchical equations of motion formalism to models with a position-dependent metal-molecule coupling. We show that the addition of a position-dependent metal-molecule coupling adds new contributions to the electronic friction and other forces, which are highly relevant for many physical processes. Our expressions for the electronic forces within the Langevin equation are valid both in and out of equilibrium and for molecular models containing strong interactions. We demonstrate the approach by applying it to different models of interest.
\end{abstract}

\maketitle

\section{Introduction}\label{sec:Introduction}
\noindent
Investigating the dynamics of molecules interacting with metal surfaces is a crucial task for understanding many phenomena in physics and chemistry. This broad research area describes a plethora of scenarios, such as the scattering of molecules off surfaces \cite{PhysRevLett.118.256001, PhysRevLett.116.217601, C8SC03955K, D2CP03312G} or chemical reactions and catalytic processes \cite{doi:10.1021/cr00077a003, PhysRevB.52.6042,PhysRevLett.89.126104}. When multiple surfaces are considered, moreover, this also includes nonequilibrium transport in molecular nanojunctions \cite{vonHippel1956,Aviram1974,Cuevas2010, Zimbovskaya2011, Bergfield2013, Thoss2018}.

However, simulating the dynamics of such systems is a challenging task. One must not only describe the interaction between the molecule and the metal surfaces accurately, but also the interaction between the electronic and vibrational degrees of freedom (DOFs) within the molecule itself. In the context of molecular nanojunctions, for example, the electronic-vibrational interaction can lead to interesting transport phenomena such as current rectification, negative differential resistance, and current induced bond rupture \cite{Galperin2007, Hrtle2011,Erpenbeck2018,Erpenbeck2020,Ke2021, Ke2023, Persson1997, Ring2020, Sabater2015}. Due to the continuum of electronic states within the metal surfaces, such electronic-vibrational interactions often result in highly nonadiabatic dynamics that requires treatment beyond the Born-Oppenheimer approximation. 

Although the most accurate approaches, such as the hierarchical equations of motion (HEOM) \cite{Tanimura1989, Schinabeck2020,Jin2008, Hrtle2013,Hrtle2018,Schinabeck2016} and the multilayer multiconfigurational time-dependent Hartree method \cite{10.1063/1.3173823,10.1063/1.3660206,10.1063/1.4965712}, treat all DOFs quantum mechanically, they can be numerically costly. This can occur especially in regimes where one needs a large vibrational basis, such as for heavy molecules or large-amplitude vibrational motion. To circumvent this problem, one can use a mixed quantum-classical method, in which vibrational DOFs within the molecule are treated classically while influenced by quantum mechanical electronic DOFs, both in the molecule and the surfaces. Beyond numerical efficiency, such treatments have the additional benefit that they often give valuable insight into the electronic forces acting on the molecular vibrations \cite{Teh2021, Teh2022}. This information can be especially insightful in the context of vibrational instabilities \cite{L2010, Todorov2010, L2011} and can be difficult to extract from a fully quantum mechanical description of the system. Within the family of such mixed quantum-classical approaches, popular methods include Ehrenfest dynamics \cite{Erpenbeck2018, Preston2022,Horsfield2004, Horsfield2004_2,Verdozzi2006, Todorov2010,Kartsev2014,Subotnik2010, Todorov2014, Cunningham2015,Bellonzi2016}, surface hopping 
\cite{10.1063/1.1675788,10.1063/1.459170,10.1063/1.4908032,10.1126/science.1179240}, and Langevin dynamics \cite{HeadGordon1995,Bode2012,L2012, Dou2016, Dou2016_2,Maurer2016, Dou2017, Dou2017_1, Dou2018, Chen2018, Chen2019, Preston2021, PhysRevB.83.115420}.

Within the Langevin dynamics approach, one first calculates the electronic friction and other forces via some quantum mechanical method, which, given the complexity of molecular models, is a nontrivial task. Many approaches to calculating electronic friction have been developed, such as using a first-principles calculation \cite{PhysRevB.94.115432}, path integrals \cite{PhysRevB.77.195316, PhysRevLett.100.176403}, scattering theory \cite{Bode2012}, nonequilibrium Green's functions (NEGFs) \cite{Dou2017}, and quantum master equations \cite{10.1063/1.4927237, 10.1063/1.4959604}. In this context, we have developed the HEOM-LD approach which combines the numerically exact HEOM method with Langevin dynamics (LD) \cite{Rudge2023,Rudge2024}.
Since the HEOM method allows for numerically exact simulation of systems with strong metal-molecule interactions, non-Markovian effects in the surfaces, and strong electron-electron or electronic-vibrational interactions within the molecule, it is a particularly promising approach for calculating electronic friction. 


So far, however, the HEOM-LD approach has incorporated the electronic-vibrational coupling only within the molecule itself, neglecting any metal-molecule coupling that depends on the position of vibrational coordinates, which can strongly influence the vibronic dynamics. Natural examples include scattering \cite{PhysRevLett.118.256001, C8SC03955K, D2CP03312G} and desorption \cite{Erpenbeck2018, Erpenbeck2020, Erpenbeck2019, PhysRevLett.116.217601} processes at metal surfaces, where the metal-molecule coupling depends sensitively on the distance to the surface. Additionally, in the context of transport through molecular nanojunctions, position-dependent metal-molecule couplings are crucial in understanding vibrational instabilities  \cite{D3NR02176A} and quantum shuttling \cite{Gorelik1998, Lai2015}. Furthermore, while there exist methods capable of calculating electronic friction with a position-dependent metal-molecule coupling \cite{Dou2016}, none of these can incorporate strong interactions within the molecule in an exact fashion. For this reason, in this paper, we extend the HEOM-LD approach to systems with a position-dependent metal-molecule coupling. 

While our approach is able to describe molecules interacting with metal surfaces in general, we specifically focus on molecular junctions for the examples that we show in this paper later on. By comparing the electronic friction calculated via HEOM to that calculated via NEGFs, which can be used for noninteracting systems, we verify the updated HEOM approach. Following this, we calculate the electronic friction of an interacting vibronic model via HEOM. Here, we demonstrate the usefulness of our extended approach since the HEOM remains numerically exact even for interacting systems. 

The paper is structured as follows. In Sec.~\ref{sec: Model}, we introduce a general model of a molecule interacting with metal surfaces. In Sec.~\ref{sec: Quantum transport theory}, we give a short overview of the HEOM approach. Following this, in Sec.~\ref{sec: Electronic friction And Langevin Dynamics}, we generalize the HEOM approach to electronic friction to systems with position-dependent metal-molecule couplings. In Sec.~\ref{sec: Results}, we apply the approach to different models of molecular nanojunctions. 

Since we have to differentiate between classically and quantum mechanically treated coordinates and momenta, we will explicitly denote the vectors of position and momentum operators as $\hat{\mathbf{x}}$ and  $\hat{\mathbf{p}}$, while their classical counterparts will be denoted by $\mathbf{x}$ and $\mathbf{p}$. Moreover, in this work, we use units where $e=\hbar=1$.

\section{Model}\label{sec: Model}
In this section, we introduce the general model considered within this work. The total setup comprises a molecule that is attached to one or more metal leads or surfaces. The entire setup is therefore described by the Hamiltonian
\begin{equation}
    H=H_{\text{mol}} +H_{\text{met}} +H_{\text{met-mol}},
\end{equation}
where $H_{\text{mol}}$ is the Hamiltonian of the molecule, $H_{\text{met}}$ is the Hamiltonian of the metal surface(s), and $H_{\text{met-mol}}$ is the metal-molecule interaction Hamiltonian. In general, the molecular Hamiltonian is given by
\begin{equation}
\begin{aligned}
        H^{}_\mathrm{mol}
    &=
     \sum_{m m'} \epsilon^{\ph}_{mm'}(\hat{\mathbf{x}})d^{\dagger}_m d^{\ph}_{m'} +\sum_{m, m'<m} U^{\ph}_{m m'}(\hat{\mathbf{x}})d^{\dagger}_m d^{\ph}_{m}d^{\dagger}_{m'} d^{\ph}_{m'} 
     \\
     &~~~+\sum_{i}\frac{\hat p_{i}^2}{2M_i}+U_{\text{\text{vib}}}(\hat{\mathbf{x}}).
\end{aligned}
\end{equation}
Here, $\hat{\mathbf{x}} = \left(\hat{x}_{1},\hat{x}_{2},\dots\right)$ and $\hat{\mathbf{p}} = \left(\hat{p}_{1},\hat{p}_{2},\dots\right)$ are vectors of the position and momentum operators of the vibrational coordinates with corresponding mass $M_i$, respectively.
Moreover,  $U_{\text{\text{vib}}}( \hat{\mathbf{x}})$ describes the potential energy surface of a reference electronic state of the molecule, e.g. the electronic ground state of the neutral molecule.
When $m = m'$, $\epsilon_{mm'}(\hat{\mathbf{x}})$ represents the energy of the $m$th electronic level of the molecule, with the coupling between states $m$ with $m'$ given by $m \neq m'$. The corresponding fermionic creation and annihilation operators for level $m$ are $d_m^{\dagger}$ and $d_{m}$, respectively. The electron-electron interaction between two electrons in state $m$ and $m'$ is considered via the matrix $U_{mm'}(\hat{\mathbf{x}})$. Although spin has not been written explicitly within $H_{\text{mol}}$, it can be included with no loss of generality to the later theory. Note that the electronic energies, hopping between electronic states, and electron-electron interactions can all depend on the vibrational coordinates, such that these terms already incorporate the electronic-vibrational interaction \cite{PhysRevB.93.115421}.

The metal surfaces or leads are modeled as reservoirs of noninteracting electrons, with Hamiltonian 
\begin{equation}
    H_{\text{met}} = \sum_{\alpha} \sum_{k} \epsilon^{\ph}_{k \alpha} c^{\dagger}_{k \alpha}c^{\ph}_{k \alpha}, 
\end{equation} 
where $\epsilon_{k \alpha}$ is the energy of state $k$ in lead or surface $\alpha$, and $c^{\dagger}_{k \alpha} $ and $c_{k \alpha}$ are the respective creation and annihilation operators. The surfaces are held at local equilibrium, with a well-defined chemical potential, $\mu_{\alpha}$, and temperature, $T$. Note that, since we exclusively investigate molecular nanojunctions in this work, we will consider two surfaces, such that $\alpha \in \{L,R\}$, where $L$ and $R$ refer to the left and right leads of the nanojunction, respectively.

The interaction between the molecule and the metal is described by the metal-molecule interaction Hamiltonian
\begin{equation}
    H_\mathrm{met-mol}
    =
    \sum_{k, \alpha, m} g_{\alpha}(\mathbf{\hat x})t_{k\alpha,m}\left( c_{k {\alpha}}^{\dagger}d^{\ph}_m +d_{m}^{\dagger}c^{\ph}_{k {\alpha}}\right),
\end{equation}
where $g_{\alpha}(\mathbf{\hat x})$ is a real-valued function describing the spatial dependency of the metal-molecule coupling, and $t_{k\alpha,m}$ describes the coupling strength between state $m$ in the molecule and state $k$ in lead $\alpha$. 
Here, we explicitly assumed that the position dependency of the metal-molecule coupling affects all states $k$ equally. 

We note that, unlike previous calculations of electronic friction using HEOM, the metal-molecule coupling depends explicitly on the vibrational coordinates $\hat{\mathbf{x}}$. This means that the level-width function of surface $\alpha$, 
\begin{equation}
    \Gamma_{\alpha,m m'} (\epsilon,\hat{\mathbf{x}}) = 2 \pi \sum_{k} g^2_\alpha(\mathbf{\hat{x}})t_{k\alpha,m}t_{k\alpha,m'} \delta(\epsilon-\epsilon_{k \alpha}).
\end{equation}
also depends on $\hat{\mathbf{x}}$. 

Initially, the molecule and the surfaces are assumed to be uncoupled, such that the total density matrix of the setup, $\rho_{\text{T}}$, factorizes
\begin{align}
    \rho_{\text{T}}(0) & = \rho_{\text{mol}}(0)\rho_{\text{met}}(0).
\end{align}
Here, $\rho_{\text{mol}}$ is the initial state of the molecule and $\rho_{\text{met}}(0)$ is the initial state of the surfaces, which is a tensor product of Gibbs' states:
\begin{equation} \label{gibbs}
    \rho_{\text{met}}(0)= \prod_{\alpha} \frac{e^{-\left(H_{\text{met},\alpha} - \mu_{\alpha}\right)/k_{\text{B}}T}}{\text{tr}_{\text{met},\alpha}\left[ e^{-\left(H_{\text{met},\alpha} -\mu_{\alpha} \right)/{k_{\text{B}}T}}\right]}.
\end{equation}

The interaction between the molecule and the metal surfaces is completely described by the correlation function 
\begin{equation}
\begin{aligned}
    C_{\alpha,mm'}^{\sigma}(\hat{\mathbf{x}},t-\tau) = \sum_{k} & g^2_{\alpha}(\mathbf{\hat x})t_{k\alpha,m}t_{k\alpha,m'} \times  \\
    & \text{tr}_{\text{met}}\left[c_{k \alpha}^{\sigma}(t) c_{k \alpha}^{\bar \sigma}(\tau)\rho_\text{met}(0) \right],
\end{aligned}
    \end{equation}
where $\sigma  = \pm$ and $\bar{\sigma} = \mp$, $c_{k \alpha}^{-}=c_{k \alpha}^{\ph}$ and $c_{k \alpha}^{+}=c_{k \alpha}^{\dag}$, and $c_{k \alpha}^{\sigma}(t) = e^{iH_{\text{met}}t}c_{k \alpha}^{\sigma}e^{-iH_{\text{met}}t}$.

The correlation and level-width functions are connected by Fourier transform via
\begin{equation}\label{bath_correlation_function}
    C_{\alpha,mm'}^{\sigma}(\hat{\mathbf{x}},t)= \frac{1}{2 \pi}\int d \epsilon ~e^{i \sigma \epsilon}\Gamma_{\alpha, mm'}(\epsilon,\hat{\mathbf{x}}) f_{\alpha}^{\sigma}(\epsilon),
\end{equation}
 where
\begin{equation}
    f_{\alpha}^{\sigma}(\epsilon)= \frac{1}{1+e^{\sigma  \left(\epsilon -\mu_\alpha \right)/k_{\text{B}}T}}
\end{equation}
is the Fermi-Dirac distribution of surface $\alpha$. The correlation functions of the metal can be expanded as a sum of exponential functions,
\begin{equation}
    C_{\alpha,mm'}^{\sigma}(t) =g^2_{\alpha}(\mathbf{\hat  x})t_{\alpha,m}t_{\alpha,m'} \sum_{\ell = 0}^{\ell_{\max}}  \eta_{\alpha,\sigma,\ell,m}e^{-\kappa_{\alpha,\sigma, \ell,m} t},
\end{equation}
which is an essential step in the HEOM approach. To calculate the coefficients $ \eta_{\alpha,\sigma,\ell,m}$ and $\kappa_{\alpha,\sigma, \ell,m}$, we decompose the Fermi-Dirac distribution via a sum-over-poles approach based on the AAA method \cite{PhysRevB.107.195429}, or a Pad\'e decomposition \cite{PhysRevB.75.035123}, which performs better than a Matsubara decomposition. For a given spectral density, one can then calculate $ \eta_{\alpha,\sigma,\ell,m}$ and $\kappa_{\alpha,\sigma, \ell,m}$ by applying residue theory to  Eq.~(\ref{bath_correlation_function}). 

Although such decomposition methods can be applied to general $\Gamma_{\alpha,m m'} (\epsilon,\hat{\mathbf{x}})$, in this work we exclusively consider the energy dependence of the level-width function to be a Lorentzian,
\begin{equation}\label{eq: gamma assumption}
    \Gamma_{\alpha,m m'} (\epsilon,\hat{\mathbf{x}}) = 2\pi t_{\alpha,m} t_{\alpha,m'} g^2_\alpha(\mathbf{\hat{x}})\frac{ W^2_\alpha}{W^2_\alpha+(\epsilon-\mu_\alpha)^{2}},
\end{equation}
where $W_\alpha$ is the bandwidth. The constant $\Gamma_{\alpha,mm'} = 2 \pi t_{\alpha,m}t_{\alpha,m'}$ reflects the maximum of the energy-dependent part of the level-width function, which is independent of the individual bath states, and will be referred to from now on as the metal-molecule coupling strength.

\section{Quantum Transport Theory}\label{sec: Quantum transport theory}

The dynamics of the entire molecule-metal system are described by the Liouville-von Neumann equation
\begin{equation}\label{von_Neumann}
    \frac{\partial}{\partial t} \rho_{\text{T}} =  -i\left[ H,\rho_{\text{T}}\right],
\end{equation}
where $\rho_{\text{T}}(t)$ is the total density matrix at time $t$, including the metal surfaces.
However, one is usually only interested in the explicit dynamics of the molecule itself, which is obtained by tracing out the metal DOFs 
\begin{equation}
     \rho_{\text{mol}} = \text{tr}_{\text{met}} \left[ \rho_{\text{T}}\right].
\end{equation}
The dynamics of the molecular density matrix is then given by
\begin{equation}\label{von_Neumann_trace}
    \frac{\partial}{\partial t} \rho_{\text{mol}} = -i \text{tr}_{\text{met}} \left(\left[ H,\rho_{\text{T}}\right]\right).
\end{equation}
Although we are only interested in the dynamics of the molecule, the interaction of the molecule with the metal still influences the dynamics of the molecule itself, which makes the solution of Eq.~(\ref{von_Neumann_trace}) challenging to obtain.

There are a variety of methods to circumvent this issue and to describe the influence of the metal on the molecule, some of which have already been mentioned in the introduction. In this work, we use the numerically exact HEOM approach based on the Feynman-Vernon influence functional. Here, we only give a short overview over the HEOM for the sake of comprehensibility of the derivations and calculations in this paper. For a more detailed description and derivation of the HEOM, we refer to Refs.~\cite{Tanimura1989,Tanimura2006,Jin2008,Ye2016,Schinabeck2016}, and the recent review of Tanimura \cite{Tanimura2020}.

The main idea of the HEOM approach is to describe the dynamics of the molecular density matrix as coupled to so-called auxiliary density operators (ADOs) of different tiers, $\rho^{(n)}_{\mathbf{j}}$. In this formalism, the zeroth tier ADO corresponds to the molecular density matrix, $\rho^{(0)} = \rho^{\ph}_{\text{mol}}$, while higher tier ADOs contain information about metal-molecule interactions and non-Markovian effects in the metal surfaces \cite{Schinabeck2016}. The ADOs are connected by a hierarchy of first-order coupled differential equations. 

Given the form of metal-molecule coupling and the factorized initial state outlined in Sec.\ \ref{sec: Model}, the equation of motion for an $n$th-tier ADO is 
\begin{equation}\label{HEOM}
\begin{aligned}
                \frac{\partial}{\partial t} \rho_{\textbf{j}}^{(n)}
                &=
                -i \left[H_\text{mol},\rho_{\textbf{j}}^{(n)}\right]- \left(\sum_{r=1}^n \kappa_{j_r} \right) \rho_{\textbf{j}}^{(n)} 
                \\
                &~~~-i\sum_{r=1}^{n}(-1)^{n-r}{\mathcal{C}}_{j_r} \rho^{(n-1)}_{\textbf{j}^-}
-i \sum_{j}\mathcal{A}^{\bar{\sigma}}_{\alpha,m} \rho^{(n+1)}_{ \textbf{j}^+}.
\end{aligned}
\end{equation}
with the coupling-up superoperator $ \mathcal{A}^{\Bar{\sigma}}_{\alpha,m}$ and the coupling-down superoperator $\mathcal{C}_{j_r}$ defined by their action on $ \rho_{\bold j}^{(n)}$,
\begin{equation}\label{coupling_up_superoperator}
     \mathcal{A}^{\Bar{\sigma}}_{\alpha,m}    \rho_{\bold j}^{(n)} = g_{\alpha}(\mathbf{\hat x})t_{\alpha,m} d^{{ \bar \sigma}}_{m} \rho_{\bold j}^{(n)}
    +(-1)^n  \rho^{(n)}_{\bold j}g_{\alpha}(\mathbf{\hat x})t_{\alpha,m} d_m^{\bar \sigma},
\end{equation}
and
\begin{equation}\label{coupling_down_superoperator}
   \mathcal{C}_{j_r} \rho_{\bold j}^{(n)}
    =
     \eta^{\ph}_{j_r} g_{\alpha}(\mathbf{\hat x})t_{\alpha,m}d^{{\sigma}}_{m}\rho_{\bold j}^{(n)}
    -(-1)^n  \rho^{(n)}_{\bold j}\eta^{*}_{\bar{j}_r} g_{\alpha}(\mathbf{\hat x})t_{\alpha,m} d_m^{\sigma}.
\end{equation}
Equations~(\ref{HEOM})-(\ref{coupling_down_superoperator}) use the superindex ${\bold j}=[{j_n,\dots,j_1}]$, $j_r=( \alpha_{j_r},\sigma_{j_r},\ell_{j_r},m_{j_r})$, with $\bar{j}_r=( \alpha_{j_r}, \bar{\sigma}_{j_r},\ell_{j_r},m_{j_r})$. Moreover, the superindex ${\bold j_{-}} = [j_n,\dots,{j_{r+1}},{j_ {r-1}},\dots,j_1]$ is formed by removing the index $j_r$ and the superindex ${\bold j_{+}} = [j,j_n,\dots,j_1]$ is formed by adding an additional index $j$.

\section{Electronic Friction and Langevin Dynamics} \label{sec: Electronic friction And Langevin Dynamics}
It is often more efficient to use mixed quantum-classical approaches, in which the vibrational DOFs of the molecule are treated classically, as influenced by quantum mechanical electronic DOFs. In this section, we will show how to obtain such a  mixed quantum-classical description. 
\subsection{Partial Wigner Transform}

We start by considering the partial Wigner transform of an operator $O$ with respect to $N$ vibrational DOFs,
\begin{equation}
\begin{aligned}
      O^{\mathcal{W}}(\textbf{x},\textbf{p})
      &=\frac{1}{(2 \pi)^{3N}} \int d\textbf{y}~ e^{i\textbf{p}\textbf{y}} \biggl \langle \textbf{x}- \frac{\textbf{y}}{2} \bigg | O \bigg |\textbf{x}+\frac{\textbf{y}}{2} \biggr \rangle,
    \end{aligned}
\end{equation}
where $\textbf{x}=(x_1,\dots, x_N)$, $\textbf{p}=(p_1,\dots, p_N)$, and the superscript $\mathcal{W}$ indicates that an operator has been Wigner-transformed \cite{Kapral1999,Wigner1932,Moyal1949}. For the sake of better readability later on, we will no longer explicitly write down the $\textbf{x}$ and $\textbf{p}$ dependency of the Wigner-transformed operators except when introducing new quantities for which the dependence is important.

The Wigner transform of the product of two operators, $O$ and $Q$, is \cite{mre1967} 
\begin{equation}\label{wigner_exponential}
\begin{aligned}
        (OQ)^{\mathcal{W}}&= O^{\mathcal{W}} e^{-i \Lambda/2} Q^{\mathcal{W}},
\end{aligned}
\end{equation}
where 
\begin{equation}
    \Lambda =\overleftarrow{\nabla}_\textbf{p}\overrightarrow{\nabla}_\textbf{x}-\overleftarrow{\nabla}_\textbf{x}\overrightarrow{\nabla}_\textbf{p},
\end{equation}
and the arrows indicate in which direction the derivatives act. A classical approximation is obtained by expanding Eq.~(\ref{wigner_exponential}) up to first order in $\Lambda$, which results in
\begin{equation}
   (OQ)^\mathcal{W} \approx O^{\mathcal{W}}Q^{\mathcal{W}}- i \big\{O^{\mathcal{W}},Q^{\mathcal{W}}\big \},
\end{equation}
with the Poisson-bracket
\begin{equation}
    \big\{ O^{\mathcal{W}},Q^{\mathcal{W}} \big\}= \sum^N_{i=1}\left(\frac{\partial O^{\mathcal{W}}}{\partial x_i} \frac{\partial Q^{\mathcal{W}}}{\partial p_i} -\frac{\partial O^{\mathcal{W}}}{\partial p_i} \frac{\partial Q^{\mathcal{W}}}{\partial  x_i} \right).
\end{equation}
Moreover, we introduce the notation
\begin{equation}
    \big \{O^{\mathcal{W}},Q^{\mathcal{W}} \big \}_\text{a} =\frac{1}{2} \left( \big\{ O^{\mathcal{W}},Q^{\mathcal{W}}\big\} - \big\{Q^{\mathcal{W}},O^{\mathcal{W}} \big \} \right).
\end{equation}
Within this classical approximation for the vibrational DOFs, the time evolution of the Wigner-transformed total density matrix, $\rho^\mathcal{W}(\mathbf{x},\mathbf{p},t)$, follows the quantum-classical Liouville equation (QCLE) \cite{Kapral1999, stock},
\begin{equation}\label{von_Neumann_trace_wigner}
    \frac{\partial}{\partial t} \rho^\mathcal{W}_{} = -i \left[ H^\mathcal{W},\rho_{\text{T}}^\mathcal{W}\right]+ \big \{ H^{\mathcal{W}},\rho_{\text{T}}^{\mathcal{W}} \big \}_\text{a}.
\end{equation}

From this, the phase-space probability density (PSPD) of the now classical variables is introduced, 
\begin{equation}\label{PSPD_definition}
    A(\mathbf{x},\mathbf{p}, t)=\text{tr}_{\text{mol}^{\text{qu}}+\text{met}} \left[ \rho^{{\mathcal{W}}}_{{\text{T}}}(\mathbf{x},\mathbf{p} , t)\right],
\end{equation}
where we have used the notation $\text{tr}_{\text{mol}^{\text{qu}}+\text{met}} \left[ \dots \right]$, which denotes the trace over the electronic DOFs in the metal and the remaining quantum DOFs in the molecule after the Wigner transformation. We note that, in the case of a constant metal-molecule coupling, Eq.~(\ref{PSPD_definition}) simplifies to the expression in Ref.~\cite{Rudge2023},
\begin{equation}\label{PSPD_old}
    A(\mathbf{x},\mathbf{p},t)=\text{tr}_{\text{mol}^{\text{qu}}} \left[ {\rho}^{\mathcal{W}}_{\text{mol}}(\mathbf{x},\mathbf{p},t)\right],
\end{equation}
because all operators depending on $\textbf{x}$ and $\textbf{p}$ are contained only in the molecular DOFs. 

The time evolution of the PSPD from Eq.~(\ref{PSPD_definition}) can be obtained by tracing out the quantum DOFs from the QCLE in Eq.~(\ref{von_Neumann_trace_wigner}), which yields 
\begin{equation} \label{dAdt_comm}
    \begin{aligned}
        \frac{\partial}{\partial t} A &=         \text{tr}_{\text{mol}^{\text{qu}}+\text{met}} \left[\big\{ H^{\mathcal{W}}_{\text{mol}}+H^{\mathcal{W}}_{\text{met-mol}},\rho^{{\mathcal{W}}}_{{\text{T}}} \big\}_\text{a}\right].
    \end{aligned}
\end{equation}
Although we have simplified the problem somewhat via the classical approximation for the vibrational DOFs, the metal-molecule interaction Hamiltonian $H^{\mathcal{W}}_{\text{met-mol}}$ still contains products of operators in the molecule and metal subspaces. Hence, performing the trace over the joint Hilbert space of molecule and metal is still a difficult task. We now show, therefore, how to express the time evolution of the PSPD in terms of the HEOM formalism.

\subsection{Wigner-Transformed HEOM}

In this section, we apply the partial Wigner transform to Eq.~(\ref{HEOM}), deriving an analogue to the quantum-classical Liouville equation in the HEOM formalism. The Wigner-transformed HEOM for the case of a position-independent metal-molecule coupling has already been introduced in Ref.\ \cite{Rudge2023}, with similar discussions for bosonic reservoirs also in Ref.\ \cite{10.1063/5.0011599}. The focus in this section, therefore, will be on how the inclusion of position-dependent $g_{\alpha}(\hat{\mathbf{x}})$ affects the partial Wigner transform with respect to the vibrational DOFs.

In particular, the coupling-up,  $\mathcal{A}^{\bar{\sigma}}_{\alpha,m}(\hat{\mathbf{x}})$, and coupling-down, $ \mathcal{C}_{j_r}(\hat{\mathbf{x}})$, superoperators now depend on $\hat{\mathbf{x}}$. Thus, applying the partial Wigner transformation to the HEOM from Eq.~(\ref{HEOM}) gives additional terms compared to  Ref.~\cite{Rudge2023} and reads
\begin{widetext}
    \begin{equation} \label{HEOM_wigner_transformed}
    \begin{aligned}
        \frac{\partial}{ \partial t} \rho_{\bold{j}}^{(n),\mathcal{W}}&= -i \left[H^{\mathcal{W}}_{\text{mol}},\rho_{\bold{j}}^{(n),\mathcal{W}}\right] +\big \{ H^{\mathcal{W}}_{\text{mol}},\rho_{\bold{j}}^{(n),\mathcal{W}} \big \}_\text{a} -\left(\sum^{n}_{r=1}  \kappa_{j_r}\right) \rho_{\bold{j}}^{(n),\mathcal{W}}
        \\
        &~~~-i\sum^{n}_{r=1}(-1)^{n-r} \left(\mathcal{C}_{j_r}^{\mathcal{W}}\rho_{\bold j^-}^{(n-1),\mathcal{W}}- \frac{i}{2}\sum^N_{i=1}     \left(\mathcal{C}_{j_r}^{\mathcal{W}}\right)'_i         \frac{\partial}{\partial p_i} \rho_{\bold j^-}^{(n-1),\mathcal{W}}\right) 
        \\ 
        &~~~-i\sum^{}_{j}\left(         \mathcal{{A}}^{\Bar{\sigma},\mathcal{W}}_{\alpha,m} \rho_{\bold{j}^+}^{(n+1),\mathcal{W}}
        -\frac{i}{2}\sum^N_{i=1}  \left(\mathcal{A}^{\Bar{\sigma}, \mathcal{W}}_{\alpha,m} \right)'_i \frac{\partial}{\partial p_i}\rho_{\bold j^+}^{(n+1),\mathcal{W}}
        \right),
    \end{aligned}
\end{equation}
\end{widetext}
where we have used the notation
\begin{equation} \label{notation_1}
    \begin{aligned}
          \left(\mathcal{A}^{\Bar{\sigma}, \mathcal{W}}_{\alpha,m} \right)'_i \frac{\partial}{\partial p_i}\rho_{\bold j}^{(n),\mathcal{W}}
        = &
       \frac{\partial}{\partial x_i} 
        g_{\alpha}(\mathbf{x})t_{\alpha,m}
        d^{{ \bar \sigma}}_{m}  \frac{\partial}{\partial p_i} \rho_{\bold j}^{(n), \mathcal{W}} - \\
        & (-1)^n  \frac{\partial}{\partial p_i} \rho_{\bold j}^{(n), \mathcal{W}} \frac{\partial}{\partial x_i} g_{\alpha}(\mathbf{x})t_{\alpha,m} d^{{ \bar \sigma}}_{m},
    \end{aligned}
\end{equation}
and
\begin{equation}  \label{notation_2}
\begin{aligned}
   \left(\mathcal{C}_{j_r}^{\mathcal{W}}\right)'_i\rho_{\bold j}^{(n),\mathcal{W}}
   = &
   \eta_{j_r}\frac{\partial}{\partial x_i}g_{\alpha}(\mathbf{x})t_{\alpha,m}d^{{  \sigma}}_{m} \frac{\partial}{\partial p_i} \rho_{\bold j}^{(n),\mathcal{W}} +  \\
   & (-1)^n  \frac{\partial}{\partial p_i} \rho_{\bold j}^{(n),\mathcal{W}}\eta^{*}_{\bar{j}_r} \frac{\partial}{\partial x_i} g_{\alpha}(\mathbf{x})t_{\alpha,m} d^{{ \bar \sigma}}_{m}
\end{aligned}
\end{equation}
If we unfold all ADOs into a vector,
\begin{equation}\label{ADO_vec}
    \text{\boldmath$\rho$}^{\mathcal{W}}=  \begin{bmatrix} \rho^{(0),\mathcal{W}}, \rho_{ j_1}^{(1),\mathcal{W}},\dots,\rho_{\bold j}^{(n_{\text{}}),\mathcal{W}},\dots
    \end{bmatrix}^{\text{T}},
\end{equation}
then the Wigner-transformed HEOM from Eq.~(\ref{HEOM_wigner_transformed}) can be written in the joint Liouville space of all ADOs,
\begin{equation}\label{drhodt_new}
    \begin{aligned}\frac{\partial}{ \partial t} \text{\boldmath$\rho$}^{\mathcal{W}}
        &=
        -\mathcal{L}^{\mathcal{W}}\text{\boldmath$\rho$}^{\mathcal{W}}
          +\big \{\big \{ H^{\mathcal{W}}_{\text{mol}},\text{\boldmath$\rho$}^{\mathcal{W}}\big \}\big \}_\text{a} 
          \\
          &~~~~- \frac{i}{2}   \sum^N_{i=1} \left(\mathcal{\Tilde{{C}}}_{\partial {x}_i}^{\mathcal{W}} +\mathcal{\Tilde{{A}}}_{\partial x_i}^{\mathcal{W}}\right)\frac{\partial}{\partial p_i}\text{\boldmath$\rho$}^{\mathcal{W}},
    \end{aligned}
\end{equation}
where we have introduced the superoperator
\begin{equation}\label{Liouvillian HEOM superoperator}
    \mathcal{L}^{\mathcal{W}}\text{\boldmath$\rho$}^{\mathcal{W}}={i \left[H^{\mathcal{W}}_{\text{mol}},\text{\boldmath$\rho$}^{\mathcal{W}} \right]+\left(\Tilde{\kappa}  -  \mathcal{\Tilde{C}}^{\mathcal{W}}  - 
         \mathcal{\Tilde{{A}}}^{\mathcal{W}} \right)\text{\boldmath$\rho$}^{\mathcal{W}}},
\end{equation}
which comprises all terms of the zeroth-order expansion of the Wigner-transformed HEOM acting on $\text{\boldmath$\rho$}^{\mathcal{W}}$. Moreover, we introduced the new superoperators $\mathcal{\tilde{C}}^{\mathcal{W}}$, $\mathcal{\tilde{A}}^{\mathcal{W}}$, and $\mathcal{\tilde{\kappa}}$, which are defined by their action on the vector of ADOs. The coupling-down superoperator in the joint Liouville space, for example, is 
\begin{equation}
    \mathcal{\tilde{C}}^{\mathcal{W}}\begin{bmatrix}
        \rho^{(0),\mathcal{W}} \\
        \rho_{j_1}^{(1),\mathcal{W}} \\
        \vdots \\
        \rho_{\bold j}^{(n),\mathcal{W}}
        \\
        \vdots
    \end{bmatrix}=\begin{bmatrix}
        0\\
        -i \mathcal{{C}}^{\mathcal{W}} \rho^{(0),\mathcal{W}}\\
        \vdots\\
        -i\sum^{n}_{r=1}(-1)^{n-r} \mathcal{{C}}^{\mathcal{W}}_{j_{r}}\rho_{\bold j^{-}}^{(n-1),\mathcal{W}}
        \\
        \vdots
    \end{bmatrix}. 
\end{equation}

The next term in  Eq.~(\ref{drhodt_new}) applies a symmetrized Poisson bracket to each Wigner-transformed ADO,
\begin{align}
    \big \{\big \{ H^{\mathcal{W}}_{\text{mol}},\text{\boldmath$\rho$}^{\mathcal{W}}\big \}\big \}_\text{a} = & \left[\big \{ H^{\mathcal{W}}_{\text{mol}},\rho_{\text{mol}}^{\mathcal{W}}\big \}_\text{a},\big \{ H^{\mathcal{W}}_{\text{mol}},\rho_{j_{1}}^{(1),\mathcal{W}}\big \}_\text{a},\dots\right]^{T},
\end{align}
and arises from the $1$st-order term in the Wigner transform of the commutator with the molecular Hamiltonian in the HEOM. In a similar manner, the Wigner transform of the coupling-up and coupling-down superoperators also yield $1$st-order terms, $\left(\mathcal{A}^{\Bar{\sigma}, \mathcal{W}}_{\alpha,m} \right)'_i$ and $\left(\mathcal{C}_{j_r}^{\mathcal{W}}\right)'_i$. In the joint Liouville space, these are collected into $\mathcal{\Tilde{{C}}}_{\partial {x}_i}^{\mathcal{W}}$ and  $\mathcal{\Tilde{{A}}}_{\partial x_i}^{\mathcal{W}}$, which are also defined by their action on the vector of ADOs. The $\mathcal{\Tilde{{C}}}_{\partial {x}_i}^{\mathcal{W}}$ superoperator, for example, is 
\begin{equation}
    \mathcal{\tilde{C}}_{\partial x_i}^{\mathcal{W}}\begin{bmatrix}
        \rho^{(0),\mathcal{W}} \\
        \rho_{j_1}^{(1),\mathcal{W}} \\
        \vdots \\
        \rho_{\bold j}^{(n),\mathcal{W}}
        \\
        \vdots
    \end{bmatrix}=\begin{bmatrix}
        0\\
        -i  \left(\mathcal{{C}}^{\mathcal{W}} \right)'_i\rho^{(0),\mathcal{W}}\\
        \vdots\\
        -i\sum^{n}_{r=1}(-1)^{n-r} \left(\mathcal{{C}}^{\mathcal{W}}_{j_{r}} \right)'_i\rho_{\bold j^{-}}^{(n-1),\mathcal{W}}
        \\
        \vdots
    \end{bmatrix}.
\end{equation}
Explicit equations for all other superoperators are given in Appendix~\ref{appendix:a}. 

\subsection{Time Evolution of the Phase-Space Probability Density and Fokker-Planck Equation}

To express the time evolution of the PSPD introduced in Eq.~(\ref{PSPD_definition}) in terms of the Wigner-transformed HEOM, we follow a similar procedure as in Ref.~\cite{Jin2008}. By comparing the time evolution of the molecular density matrix, governed by the Wigner-transformed Liouville equation, and the time evolution of the zeroth tier ADO in  the Wigner-transformed HEOM from Eq.~(\ref{HEOM_wigner_transformed}), we show in Appendix~\ref{appendix:b}, that the time evolution of $A(\mathbf{x},\mathbf{p},t)$ in the HEOM formalism can be expressed as 
\begin{equation}\label{PSPD_final_vec}
    \begin{aligned}
                    \frac{\partial A }{\partial t} = \:\: & \text{tr}_{\text{mol}^{\text{qu}}} \left[ \big \{\big \{{H}^{\mathcal{W}}_{\text{mol}}, \text{\boldmath$\rho$}^{\mathcal{W}} \big \}\big \}_\text{a}\right] +  \\
                    & \text{tr}_{\text{mol}^{\text{qu}}}  \left[\sum^N_{i=1} {\mathcal{H}}_{\text{met-mol},i}^{\mathcal{W}} \frac{\partial \text{\boldmath$\rho$}^{\mathcal{W}}  }{\partial p_i}\right]. 
    \end{aligned}
\end{equation}
Here, we have introduced the new superoperator ${{\mathcal{H}}^{\mathcal{W}}_{\text{met-mol},i}}$, which is defined via its action on the vector of all ADOs,
\begin{equation} \label{h_sb_definition}
    {{\mathcal{H}}^{\mathcal{W}}_{\text{met-mol},i}}     \text{\boldmath$\rho$}^{\mathcal{W}}
    =
    \begin{bmatrix}  
     \sum\limits_{j}     \frac{\partial g_{\alpha}}{\partial x_i} t_{\alpha,m}  d_m^{\bar \sigma}\rho_{j}^{(1),\mathcal{W}}
    \\ 
    0 
    \\
    0
    \\
    \vdots 
    \end{bmatrix},
\end{equation}
and where the trace now selects and traces over only the $0$th-tier
\begin{align}
    \text{tr}_{\text{mol}^{\text{qu}}}  \left[\text{\boldmath$\rho$}^{\mathcal{W}}\right] & = \text{tr}_{\text{mol}^{\text{qu}}}  \left[\rho^{(0),\mathcal{W}}\right].
\end{align}

Although we now have an expression for time evolution of the PSPD in terms of the HEOM formalism, it is still in the form of a quantum-classical Liouville equation. In order for electronic friction to appear, we need to further apply a limit of weak nonadiabaticity, which we do by assuming a timescale separation between fast electronic relaxation and slow vibrational dynamics. In the following, we explore the effects of such an assumption. 

First, we write the vector of all ADOs with the ansatz \cite{PhysRevLett.119.046001} 
\begin{equation}\label{rho_ss}
  \text{\boldmath$\rho$}^{\mathcal{W}} (\mathbf{x},\mathbf{p},t)= A(\mathbf{x},\mathbf{p},t) \text{\boldmath$\sigma$}^{\mathcal{W}}_{\text{ss}}(\mathbf{x}) +\text{\boldmath$B$}(\mathbf{x},\mathbf{p},t),
\end{equation}
where $\text{\boldmath$\sigma$}_{\text{ss}}^{\mathcal{W}}(\mathbf{x})$ is the electronic steady state frozen at one vibrational frame:
\begin{equation}\label{Liouvillian}
     \mathcal{L}^{\mathcal{W}}\text{\boldmath$\sigma$}_{\text{ss}}^{\mathcal{W}} =0.
\end{equation}
One can view this term as the electronic steady state in the purely adiabatic limit of Eq.~(\ref{drhodt_new}). The contribution $\text{\boldmath$B$}$ simply contains the difference between the exact vibronic dynamics and this adiabatic limit.

The key step is to recognize that, under the timescale separation assumption, $\text{\boldmath$B$}$ provides only a small nonadiabatic correction. By following a similar derivation as in Ref.~\cite{Rudge2023}, we show in the Appendix~\ref{appendix:c} that, under this assumption and by combining  Eq.~(\ref{rho_ss}) with  Eq.~(\ref{PSPD_final_vec}), the time evolution of $A$ in Eq.~(\ref{PSPD_final_vec}) is given by the Fokker-Planck equation
\begin{equation}\label{fokker_planck_equation}
    \begin{aligned}
          \frac{\partial}{\partial t} A 
          &=
          -\sum_i\frac{p_i}{m_i} \frac{\partial A}{ \partial x_i} +\sum_i F^{\text{ad}}_i(\mathbf{x}) \frac{\partial A}{ \partial p_i}  + 
          \\
          &~~~ \sum_{ij}\gamma_{ij}(\mathbf{x}) \frac{\partial }{\partial p_i} \left(\frac{p_j}{m_j} A \right) +\sum_{ij}D_{ij}(\mathbf{x})\frac{\partial^2 A}{\partial p_i \partial p_j}.
    \end{aligned}
\end{equation}
The dynamics of  Eq.~(\ref{fokker_planck_equation}) are governed by the adiabatic contribution to the mean force $F^{\text{ad}}_i(\mathbf{x})$, the electronic friction tensor, $\gamma_{ij}(\mathbf{x})$, the correlation function of the stochastic force, $D_{ij}(\mathbf{x})$. Although  Eq.~(\ref{fokker_planck_equation}) has the same form as was derived in Ref.~\cite{Rudge2023}, the inclusion of a position dependence in the metal-molecule coupling adds additional terms to these electronic forces. 

The adiabatic contribution to the mean force, for example, is now 
\begin{equation}\label{mean_force}
    \begin{aligned}
        F^{\text{ad}}_i(\mathbf{x})&=F^{\text{ad}}_{i,{\text{mol}}}(\mathbf{x})+F^{\text{ad}}_{i,{\text{met-mol}}}(\mathbf{x}),
    \end{aligned}
\end{equation}
where there are contributions originating not only from the molecular Hamiltonian, but also the metal-molecule coupling:
\begin{equation}\label{mean_force_mol}
    \begin{aligned}
        F^{\text{ad}}_{i,\text{mol}}(\mathbf{x})&=-\text{tr}_{\text{mol}^{\text{qu}}}   \Bigg[ \Bigg. \frac{ \partial H^{\mathcal{W}}_{\text{mol}}}{\partial x_i} 
        \text{\boldmath$\sigma$}_{\text{ss}}^{\mathcal{W}}    \Bigg], 
    \end{aligned}
\end{equation}
\begin{equation}\label{mean_force_metal_mol}
    \begin{aligned}
        F^{\text{ad}}_{i,\text{met-mol}}(\mathbf{x})&=-\text{tr}_{\text{mol}^{\text{qu}}}   \Bigg[ \Bigg.       \mathcal{H}_{\text{met-mol},i}^{\mathcal{W}}       \text{\boldmath$\sigma$}_{\text{ss}}^{\mathcal{W}}    \Bigg].
    \end{aligned}
\end{equation}
Similarly, the electronic friction reads
\begin{equation}\label{friction_coef}
    \begin{aligned}
        \gamma_{ij}(\mathbf{x}) &= \gamma_{ij,\text{mol}}(\mathbf{x})+ \gamma_{ij,\text{met-mol}}(\mathbf{x}),
    \end{aligned}
\end{equation}
where
\begin{equation}\label{friction_coef_mol}
    \begin{aligned}
        \gamma_{ij,\text{mol}}(\mathbf{x}) &=-\text{tr}_{\text{mol}^{\text{qu}}}  \left[ {\frac{ \partial H^{\mathcal{W}}_{\text{mol}}}{\partial x_i} }_{}\left(\mathcal{L}^{\mathcal{W}}\right)^{-1}\frac{\partial \text{\boldmath$\sigma$}_{\text{ss}}^{\mathcal{W}}}{\partial x_j}\right],
    \end{aligned}
\end{equation}
\begin{equation}\label{friction_coef_metal_mol}
    \begin{aligned}
        \gamma_{ij,\text{met-mol}}(\mathbf{x}) &=-\text{tr}_{\text{mol}^{\text{qu}}}  \left[ {{\mathcal{H}}_{\text{met-mol},i}^{\mathcal{W}}} \left(\mathcal{L}^{\mathcal{W}}\right)^{-1}\frac{\partial \text{\boldmath$\sigma$}_{\text{ss}}^{\mathcal{W}}}{\partial x_j}\right].
    \end{aligned}
\end{equation}
For the correlation function of the random force we obtain
\begin{equation}\label{diffusion_coef}
    \begin{aligned}
        D_{ij}(\mathbf{x})&=\text{tr}_{\text{mol}^{\text{qu}}}   \Bigg[ \Bigg. \left(\frac{ \partial H^{\mathcal{W}}_{\text{mol}}}{\partial x_i} 
        +\mathcal{H}_{\text{met-mol},i}^{\mathcal{W}} \right)
         \left(\mathcal{L}^{\mathcal{W}}\right)^{-1} 
         \\
        & ~~~~\times \Bigg( \Bigg.   
        \left( - \frac{i}{2}    \left(\mathcal{\Tilde{{C}}}_{\partial {x}_j}^{\mathcal{W}} +\mathcal{\Tilde{{A}}}_{\partial x_j}^{\mathcal{W}}\right)\text{\boldmath$\sigma$}_{\text{ss}}^{\mathcal{W}}\right)         
          \\
          &~~~~+\frac{1}{2}\left( \frac{\partial H^{\mathcal{W}}_{\text{mol}}}{\partial x_j}\text{\boldmath$\sigma$}_{\text{ss}}^{\mathcal{W}}-\text{\boldmath$\sigma$}_{\text{ss}}^{\mathcal{W}} \frac{\partial H^{\mathcal{W}}_{\text{mol}}}{\partial x_j}\right)   +  F^{\text{ad}}_j(\mathbf{x})\text{\boldmath$\sigma$}_{\text{ss}}^{\mathcal{W}} \Bigg) \Bigg. \Bigg] \Bigg. .
    \end{aligned}
\end{equation}
In  Eqs.~(\ref{friction_coef_mol}-\ref{diffusion_coef}), $\left(\mathcal{L}^{\mathcal{W}}\right)^{-1} $ refers to a formal inversion of  Eq.~(\ref{Liouvillian HEOM superoperator}).

The Langevin equation corresponding to the Fokker-Planck equation in Eq.~(\ref{fokker_planck_equation}) is given by 
\begin{equation}\label{langevin_equation}
    m_i \Ddot{x}_{i} = F^{\text{ad}}_i(\mathbf{x})+ \sum_j\gamma_{ij}(\mathbf{x}) \dot{x}_{i} +  f_i(t) 
    \end{equation}
where  $f_i(t)$ is a Gaussian random force with white noise
\begin{equation}
    \langle  f_i(t)f_j(t')\rangle = 2D_{ij}(\mathbf{x}) \delta(t-t').
\end{equation}

\section{Results}\label{sec: Results}

In this section, we apply the extended HEOM-LD formalism introduced in the previous section to several example scenarios, investigating how the inclusion of a position-dependent metal-molecule coupling affects the electronic forces entering the Langevin dynamics. Although the extended HEOM-LD approach is applicable to the general problem of molecules interacting with metal surfaces, we will consider the specific scenario of charge transport through a molecular nanojunction, in which we have two metal surfaces, the left, $\alpha = \text{L}$, and right, $\alpha = \text{R}$, leads. These can either be held at equilibrium, such that the chemical potentials and temperatures of both leads are equal, or driven out of equilibrium via a voltage bias, $\Phi = \mu_{\text{L}} - \mu_{\text{R}}$, applied symmetrically around the Fermi level, which is always set to zero, $\varepsilon_{\text{F}} = 0$, such that $\mu_{\text{L}} = -\mu_{\text{R}} = e\Phi/2$.

The first scenario is nonequilibrium transport in a simple vibronic model, in which the molecule contains a single electronic level linearly coupled to a single harmonic vibrational mode. The second system introduces a strong electronic-vibrational interaction, such that the molecular model contains a single electronic level coupled linearly to both a low-frequency mode, which is treated classically via Langevin dynamics, and a high-frequency mode, which must be treated quantum mechanically alongside the electronic DOFs within the HEOM. For both systems, we will first investigate adiabatic and nonadiabatic effects at equilibrium, before discussing the nonequilibrium situation.



\subsection{Simple Vibronic Model}\label{subsec:verify}

First, we consider a single electronic level linearly coupled to a harmonic vibrational mode, with molecular Hamiltonian  
\begin{equation}
     H^{}_{\text{mol}}
     =\left(\epsilon_0 + \lambda   \hat x\right)d^\dagger d +\frac{\omega}{2}\left( \hat x^2 +  \hat p^2 \right),
\end{equation}
 where we have implicitly used dimensionless coordinates, $\hat x\rightarrow  \hat x\sqrt{m \omega}$ and $ \hat p \rightarrow \hat p/\sqrt{m\omega}$. Moreover, $\epsilon_0$ is the energy of the electronic level without coupling to the vibrational mode, $\lambda$ is the electronic-vibrational coupling strength, and $\omega$ is the frequency of the vibrational mode. We can also rewrite the molecular Hamiltonian as 
 \begin{equation}
     H^{}_{\text{mol}}
     =U_\text{g}(\hat x) dd^{\dag} + U_\text{c}(\hat x)d^{\dag}d + \frac{\omega}{2} \hat p^{2},
 \end{equation}
 which allows us to identify the potential energy surfaces of the neutral, $  U_\text{g}(\hat x)=\frac{1}{2}\omega \hat x^2$, and charged, $U_\text{c}(\hat x)=U_\text{g}(\hat x )+\lambda \hat x +\epsilon_0$, states of the molecule, respectively.
 
 Once the vibrational coordinate is treated classically, the electronic part of this Hamiltonian is noninteracting. Therefore, the electronic friction and other forces can be calculated exactly not only within the HEOM approach, but also using NEGFs \cite{Dou2016}. Following Ref. \cite{Dou2016} and rewriting the electronic friction in our notation, one obtains the following formula at equilibrium in the wideband limit: 

\begin{equation}\label{friction_negfs_dou}
     \begin{aligned}
     \gamma(x) = - \frac{1}{4\pi}\int d \epsilon~ & \left( \frac{\partial h}{\partial x}+\frac{\epsilon-h(x)}{\left(g(x)\right)^2}\frac{\partial \left(g\right)^2}{\partial x}\right)^2 \times
     \\
    &  A^2(\epsilon,x)\frac{\partial f(\epsilon)}{\partial \epsilon},
 \end{aligned}
\end{equation} 
where $h(x) = U_{\text{c}}(x) - U_{\text{g}}(x) = \epsilon_{0} + \lambda x$ and 
\begin{equation}
    A(\epsilon,x)= \frac{\Gamma \cdot \left(g(x)\right)^2 }{\left(\epsilon- h(x)\right)^2+\left(\Gamma \cdot\left( g(x) \right)^2 /2\right)^2}
\end{equation}
is the spectral function. Out of equilibrium, Eq.~(\ref{friction_negfs_dou}) can be generalized as in Ref.~\cite{10.1063/5.0153000}.
 In the following, we will not only investigate the electronic forces of this model, but also verify the extended HEOM-LD approach by comparing the electronic friction as calculated from Eq.~(\ref{friction_negfs_dou}) to the expression in  Eq.~(\ref{friction_coef}). To mimic the wideband approximation, we choose a bandwidth of $W_\alpha=30~\text{eV}$ for all HEOM calculations.

 To help identify the specific effect of the position-dependent metal-molecule coupling, we first investigate the electronic forces of a  constant coupling, $g_\alpha(\hat x) = 1$. Although similar systems have already been analyzed previously, for the sake of self-completeness, we will introduce the main features of the electronic forces, starting in the equilibrium situation, $\Phi = 0\text{V}$.

\begin{center}
 \begin{figure} 
     \includegraphics[width=\columnwidth]{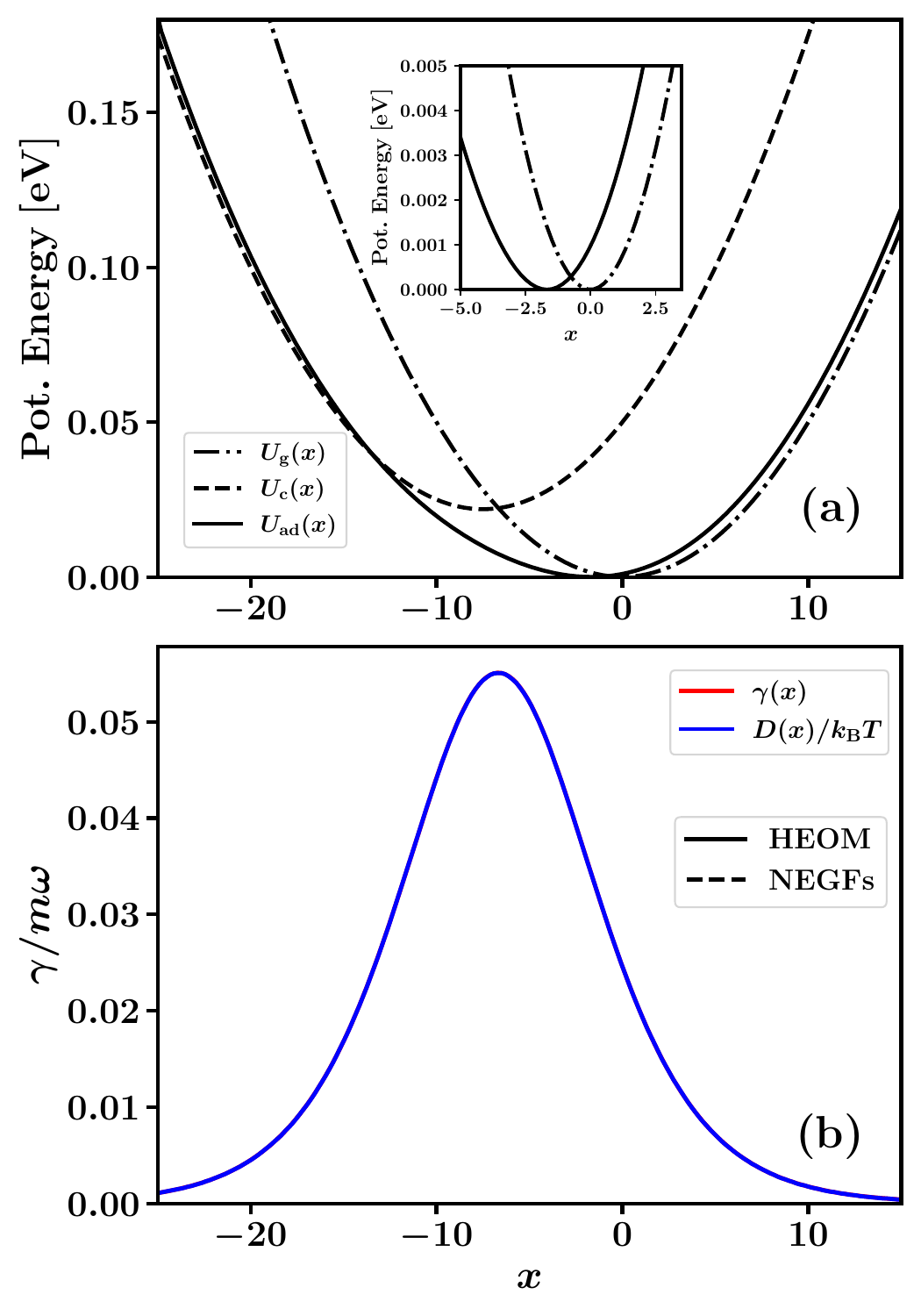}
     \centering     \phantomsubfloat{\label{fig:potentials_cl_no_x_coupling_potentials}}     \phantomsubfloat{\label{fig:potentials_cl_no_x_coupling_friction}}
     \vspace{-2\baselineskip}
     \caption{Potential energy surfaces for the vibronic model (a), and friction and correlation function calculated with HEOM and NEGFs (b). The parameters are $\epsilon_0=~0.05~\text{eV}$, $\lambda= 0.0075~\text{eV} $, $\omega= 0.001~\text{eV}$, and $\Gamma_\alpha=0.0049~\text{eV}$. The temperature of the metal is $T=300~K$.
     \label{fig:potentials_cl_no_x_coupling}}
 \end{figure}    
 \end{center}

 Fig.~\ref{fig:potentials_cl_no_x_coupling_potentials} shows the potential energy surfaces corresponding to the two diabatic potential energy surfaces in the molecular Hamiltonian, as well as the adiabatic potential energy surface calculated from the adiabatic contribution to the mean force,
 \begin{equation}
  U_{\text{ad}}(x)=\int^{x}_{x_{0}} dx' ~F^{\text{ad}}(x').
 \end{equation}
 The diabatic potential energy surfaces of the neutral and charged state cross at 
 \begin{equation}
     x_{\varepsilon_\text{F}}=-(\epsilon_0-{\varepsilon_\text{F} })/{\lambda},
 \end{equation}
 which corresponds to the coordinate at which the electronic energy equals the Fermi energy of the leads. In the vicinity of $ x_{\varepsilon_\text{F}}$, the adiabatic potential shows a typical avoided crossing, which is due to the nonadiabatic coupling between the charged and uncharged diabats via the interaction with the metal. For large negative $x$ values, the adiabatic potential follows the diabat of the charged state, since the electronic level at these positions is so far below $\varepsilon_\text{F}$ that it is mostly occupied. In contrast, for large positive $x$ values, the energy level is so far above $\varepsilon_\text{F}$ that the electronic level is mostly unoccupied and the adiabatic potential follows the neutral state diabat.

To investigate the nonadiabatic effects, we turn to Fig.~\ref{fig:potentials_cl_no_x_coupling_friction}, which shows the electronic friction for this system calculated via NEGFs as well as the HEOM-LD approach. The friction shows a peak at $ x_{\varepsilon_\text{F}}$, as the friction represents a nonadiabatic correction to the adiabatic force, and it is exactly at this position that there is an avoided crossing in $U_{\text{ad}}(x)$. In particular, the friction describes relaxation of the vibrational mode via coupling to electron-hole pairs (EHPs), in that electrons may transfer from a lead to the molecule, absorb energy from the vibrational mode, and transfer back to a higher energetic state in the lead. At equilibrium, such processes are resonant exactly at the position where the electronic level crosses the chemical potential, $ x_{\varepsilon_\text{F}}$. 

 Moreover, Fig.~\ref{fig:potentials_cl_no_x_coupling_friction} shows that the friction and the correlation function of the stochastic force not only match between the HEOM and NEGFs approaches, but that they also satisfy the fluctuation-dissipation theorem \cite{Einstein1905}:
 \begin{equation}\label{fluctuation_dissipation_formula}
     D(x)=k_{\text{B}}T \gamma(x).
 \end{equation}
 This reflects that the relaxation of vibrational modes via coupling to EHPs is balanced by the reverse process at equilibrium. 

\begin{figure}[]    
         \centering
         \includegraphics[trim=10. .0 12 .0, clip, scale=.54]{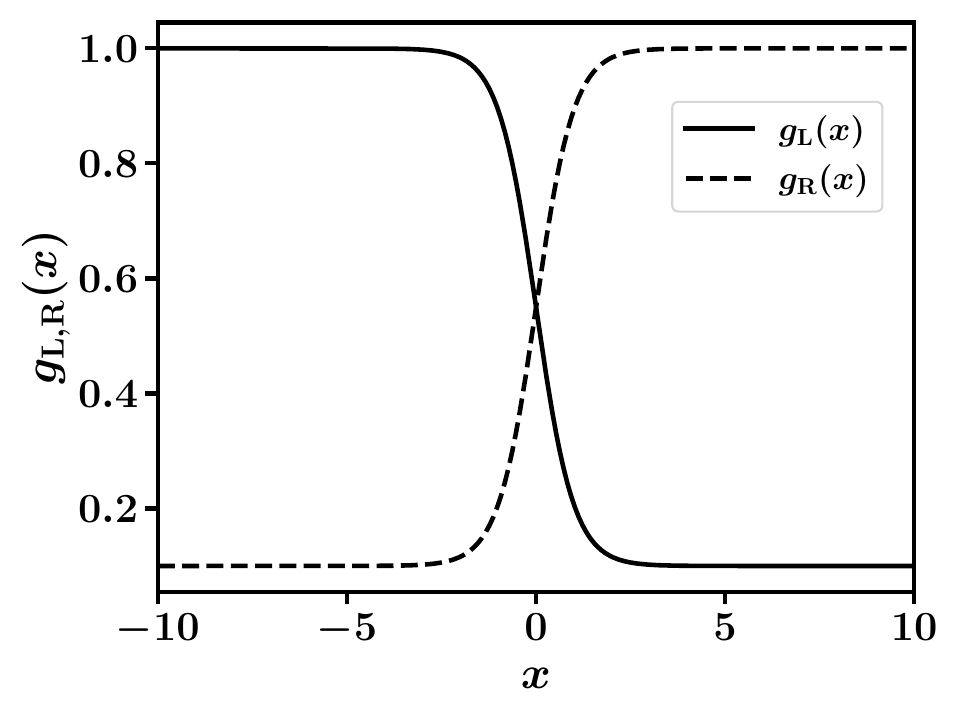}
         \caption{Spatial part of the metal-molecule couplings $g_{\text{L}/ \text{R}}(x)$ for the vibronic model in  Eq.~(\ref{mm_coupling_andre}). The parameters entering $g_{\text{L}/ \text{R}}(x)$ are $q=0.1$, and $ a =1$.}            
         \label{fig:ml_coupling}
 \end{figure}

Next, we add a position-dependent metal-molecule coupling to the model. Although one can obtain such couplings from a first-principles calculation, here, we choose a simple form of the coupling that is designed to reproduce the basic features one would expect in a molecular nanojunction. In particular, if the vibrational coordinate represents center-of-mass motion between the two leads, then the coupling $g_{\alpha}(\hat x)$ increases and decreases as the molecule approaches and leaves lead $\alpha$, respectively. The spatial part of the metal-molecule coupling, therefore, takes the form 
 \begin{equation}\label{mm_coupling_andre}
                 g_{\text{L}}(\hat x)
     =
      \left[ \frac{1-q}{2} \left( 1-\text{tanh} \left( \frac{\hat x}{ a}\right)+q \right)\right],
 \end{equation}
 and
 \begin{equation}
                 g_{\text{R}}(\hat x)
     =
     -g_{\text{L}}(\hat x)+g_\text{c},
 \end{equation}
 where we choose $g_\text{c}$ so that $ g_{\text{L}}(\pm \infty)= g_{\text{R}}(\mp \infty)$. Similar metal-molecule couplings have been used in Refs.~\cite{Erpenbeck2018,Erpenbeck2019,Erpenbeck2020,Ke2021}, and recently in Ref.~\cite{D3NR02176A}. The position dependency of the metal-molecule coupling is shown in Fig.~\ref{fig:ml_coupling}. To compare with the case of a constant metal-molecule coupling, we set the metal-molecule-coupling strength to ${\Gamma}_\alpha = 2\pi {t}_{\alpha}^{2} = 0.01~\text{eV}$, since now for $\hat x \rightarrow \pm \infty $ the molecule is essentially only coupled to one surface. 

\begin{center}
 \begin{figure}
     \includegraphics[trim=10. .0 0 .0, clip, scale=.51]{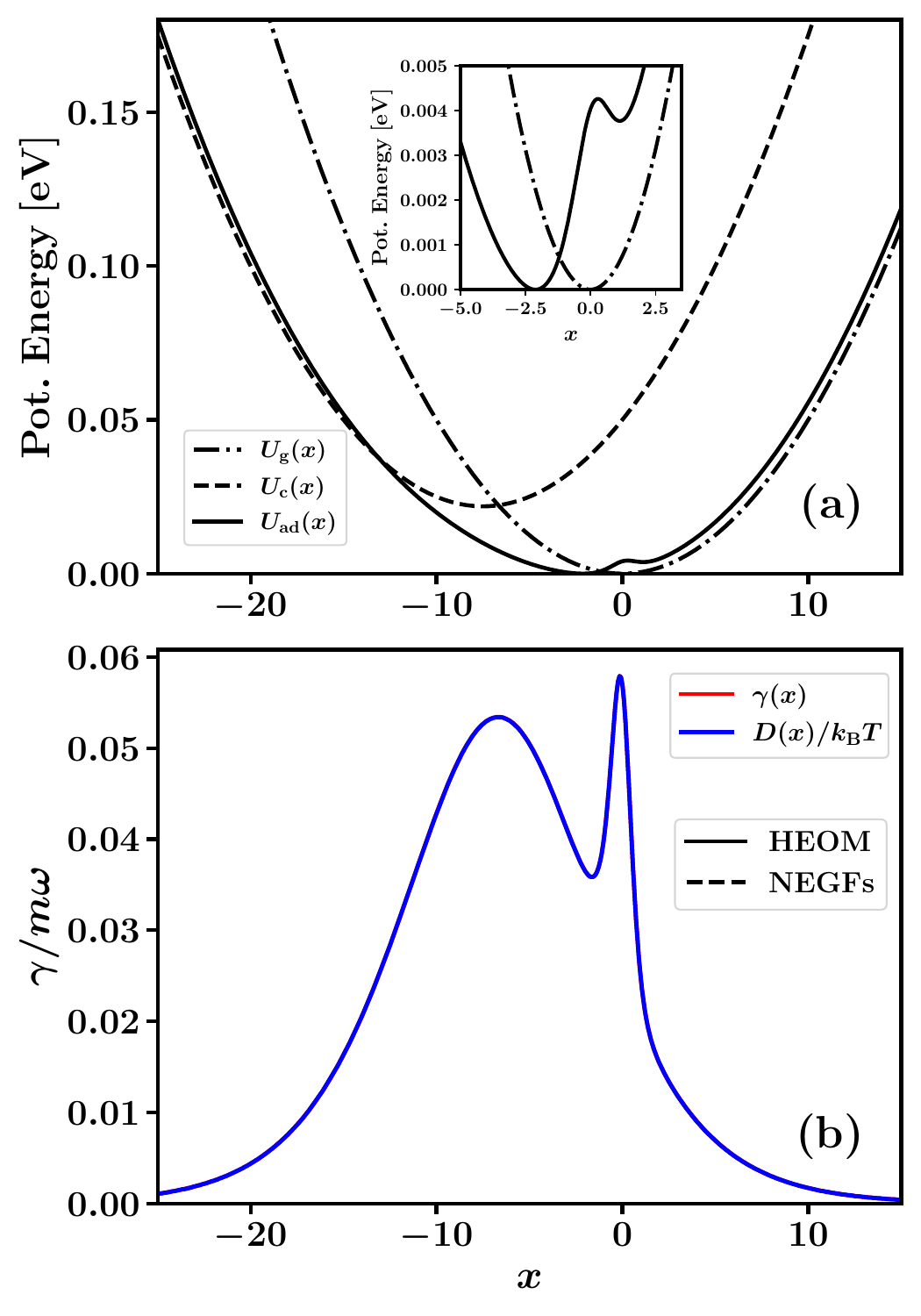}
     \centering     \phantomsubfloat{\label{fig:potentials_cl_potentials}}    \phantomsubfloat{\label{fig:potentials_cl_friction}}
     \vspace{-2\baselineskip}
     \caption{Potential energy surfaces for the system with the position-dependent metal-molecule coupling from Fig.~(\ref{fig:ml_coupling}) (a), and friction and correlation function calculated with HEOM and NEGFs (b). Apart from setting $\Gamma_\alpha=0.01~\text{eV}$, all parameters are the same as in the previous model from Fig.~(\ref{fig:potentials_cl_no_x_coupling}). 
     }    
     \label{fig:potentials_cl}
 \end{figure}    
 \end{center}

 Fig.~\ref{fig:potentials_cl_potentials} shows the potential energy surfaces for the vibronic model with the position-dependent metal-molecule coupling from  Eq.~(\ref{mm_coupling_andre}). As in the constant coupling case, $U_{\text{ad}}(x)$ displays an avoided crossing at $x_{\varepsilon_\text{F}}$. However, in the vicinity of $x = 0$, two minima appear, which originate from $g_{\alpha}(x)$. First, we note that the contribution to the adiabatic force arising from the metal-molecule coupling, $F^{\text{ad}}_{i,\text{met-mol}}(x)$, is proportional to the spatial derivative of the metal-molecule couplings, $\frac{\partial g_{\alpha}}{\partial x}$.

 At $x=0$, both $\frac{\partial g_{L}}{\partial x}$ and $\frac{\partial g_{R}}{\partial x}$ reach their extrema. Since this is close to the point at which the energy level crosses the Fermi level of the leads, any change in the vibrational coordinate at this point is accompanied by a strong change in the electronic occupancy. Thus, in the vicinity of $x=0$, the contribution to the adiabatic force from the metal-molecule coupling is maximized. Hence, the classical vibrational DOF will move away from these positions in either positive or negative direction, to positions where the forces acting on it are minimized, which is why there are two minima close to $x = 0$. 

\begin{figure}[t]    
         \centering
         \includegraphics[trim=15. .0 9 .0, clip, scale=.54]{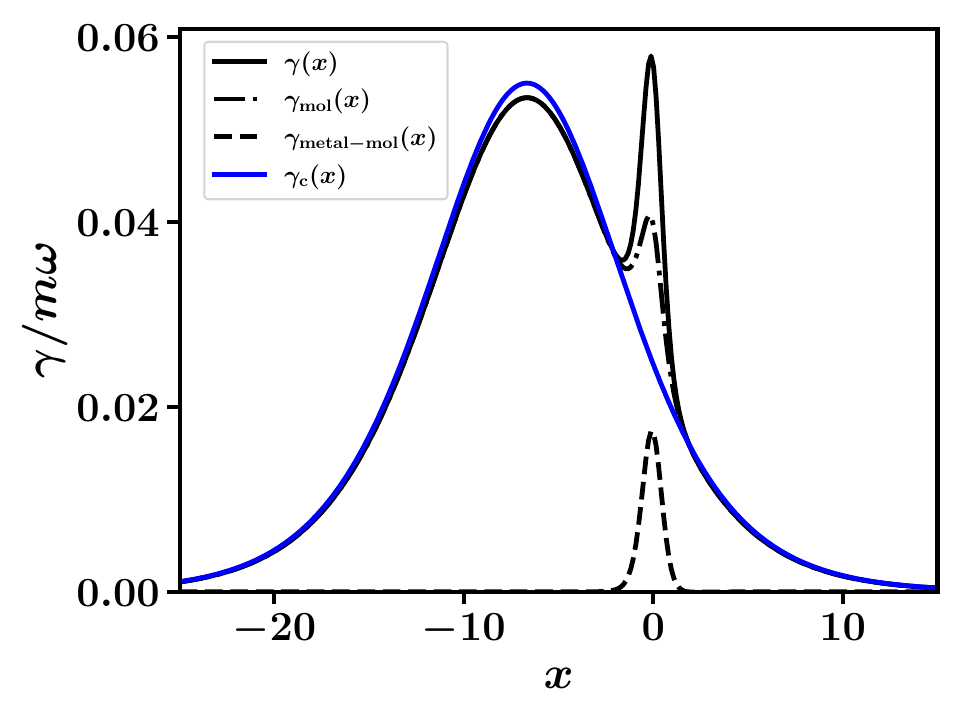}
         \caption{Contributions to the friction from the molecular Hamiltonian, $\gamma_{\text{mol}}(x)$, and the metal-molecule coupling, $\gamma_{\text{met-mol}}(x)$, as well as the friction for a constant coupling, $\gamma_\text{c}(x)$, from the previous model in Fig.~\ref{fig:potentials_cl_no_x_coupling_friction}. }            
         \label{fig:friction_contributions}
 \end{figure}

 This contribution also manifests itself in the electronic friction and correlation function, shown in Figs.~\ref{fig:potentials_cl_friction} and \ref{fig:friction_contributions}. First, we note in Fig.~\ref{fig:potentials_cl_friction} that the electronic forces calculated from the HEOM and NEGFs methods still match exactly, even with the addition of a position-dependent metal-molecule coupling. We also note that the fluctuation-dissipation theorem is still satisfied because $\Phi = 0\text{ V}$. To understand the behavior of these forces, we turn to Fig.~\ref{fig:friction_contributions}, which plots the full friction, $\gamma(x)$, as well as the contributions from the molecular Hamiltonian, $\gamma_\text{mol}(x)$, the metal-molecule interaction, $\gamma_\text{met-mol}(x)$, and the friction for a constant metal-molecule coupling from the previous model, $\gamma_\text{c}(x)$. 

 As in the constant coupling case, the full electronic friction also displays a peak at $ x_{\varepsilon_\text{F}}$, which originates from the same EHP creation processes that were discussed in Fig.~\ref{fig:potentials_cl_no_x_coupling_friction}. However, there is now an additional side peak at a point that we will name $x_{\text{mm}} $, which arises directly from the position dependency of the metal-molecule coupling. To understand the origin of this side peak, we consider the individual contributions to the total friction. First, we note that $\gamma_\text{mol}(x)$ also contains a side peak at $x_{\text{mm}}$ and does not just reproduce the friction in the constant coupling case. Although $\gamma_\text{mol}(x)$ does not depend explicitly on the metal-molecule coupling, it depends on it implicitly via $\frac{\partial \sigma^{(0),\mathcal{W}}_{\text{ss}}}{\partial x}$, which is the origin of this side peak. At position $x_{\text{mm}}$, the metal-molecule couplings are changing rapidly and the molecular electronic energy is close to the Fermi level, such that $\sigma^{(0),\mathcal{W}}_{\text{ss}}$ is also changing rapidly. When one includes the $\gamma_\text{met-mol}(x)$ contribution, which has a peak at $x_{\text{mm}}$, we recover the total friction. From a physical point of view, electronic friction represents a nonadiabatic vibrational relaxation process via coupling to EHP creation. Such processes are dominant at points where $\frac{\partial \sigma^{(0),\mathcal{W}}_{\text{ss}}}{\partial x}$ and $\frac{\partial g_{\alpha}}{\partial x}$ are large, as these are points in space at which electrons can easily jump back and forth between metal and molecule.

This discussion also helps to connect the HEOM-LD expression for the electronic friction to previous expressions, such as that shown for noninteracting electronic systems from NEGFs. In  Eq.~(\ref{friction_negfs_dou}), for example, the friction comprises four terms: two that contain only $\frac{\partial g_{\alpha}}{\partial x}$ or $\frac{\partial H^\mathcal{W}_{\text{mol}}}{\partial x}$, and two cross terms. In the HEOM-LD expression, although it appears there are only two components to the electronic friction, these cross terms are contained within $\frac{\partial \sigma^{(0),\mathcal{W}}_{\text{ss}}}{\partial x}$ and $\frac{\partial \sigma^{(1),\mathcal{W}}_{\text{ss},j}}{\partial x}$. 

\begin{figure}[t]    
         \includegraphics[trim=15. .0 9 .0, clip, scale=.51]{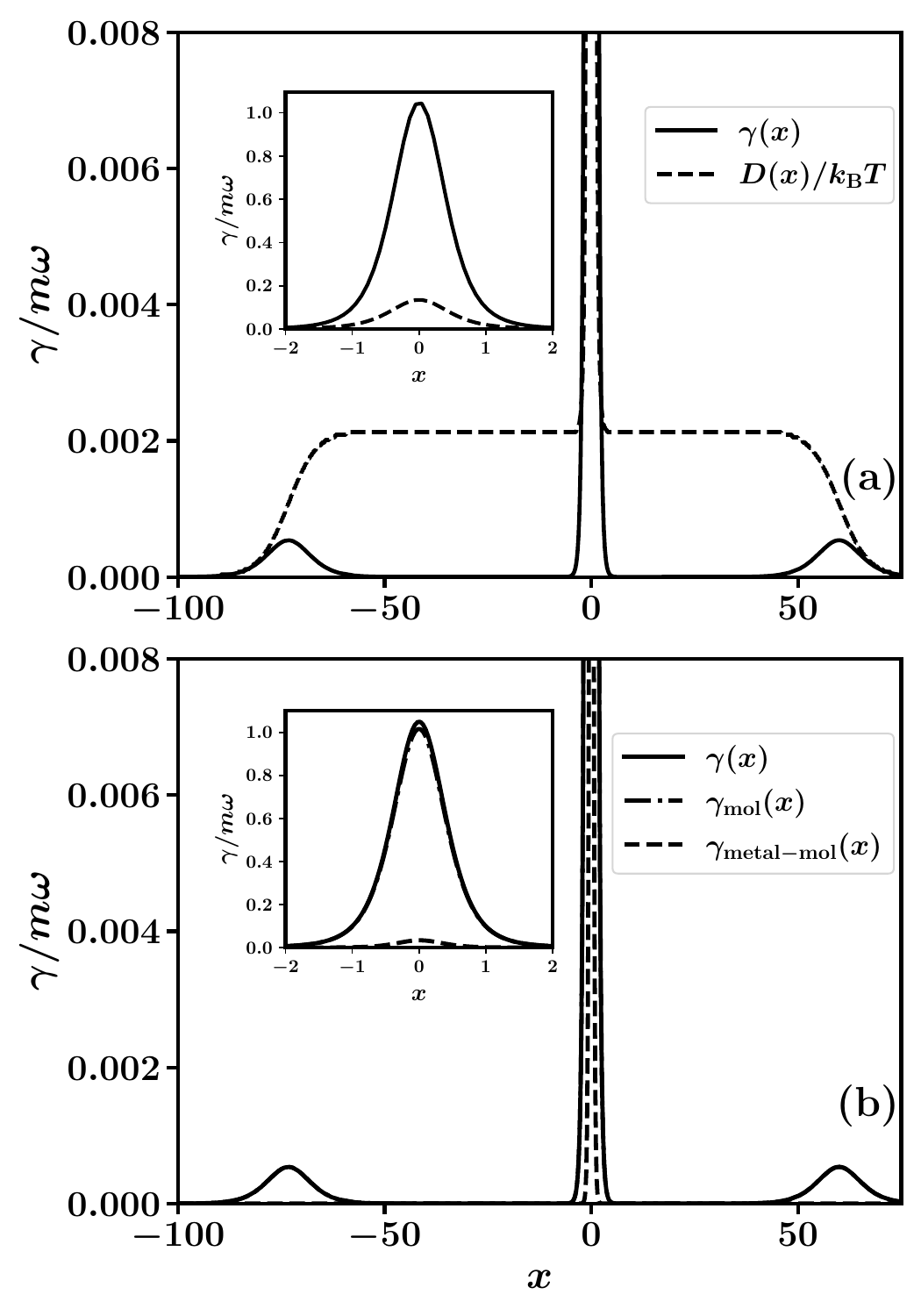}
              \centering     \phantomsubfloat{\label{fig:friction_heom_negf_noneq_friction_correlationfunction}}     \phantomsubfloat{\label{fig:friction_heom_negf_noneq_contribution}}
         \vspace{-2\baselineskip}
         \caption{Friction and correlation function out of equilibrium (a), and different contributions to the friction (b). Apart from setting $\Phi=1\text{V}$, the parameters are the same as in Fig.~\ref{fig:potentials_cl}.}            
         \label{fig:friction_heom_negf_noneq}
 \end{figure}

 Next, we drive the nanojunction out of equilibrium by applying a bias voltage, $\Phi=1~\text{ V}$. The corresponding electronic friction and correlation function of the stochastic force are shown in Fig.~\ref{fig:friction_heom_negf_noneq_friction_correlationfunction}.
 Out of equilibrium, EHP creation and annihilation processes are resonant when the electronic energy crosses $\mu_{\alpha}$. Consequently, the main equilibrium friction peak originating from $\frac{\partial H^\mathcal{W}_{\text{mol}}}{\partial x}$ splits into two peaks at positions 
 \begin{equation}
     x_{\mu_\alpha} = -(\epsilon_0-\varepsilon_{\text{F}}-\mu_\alpha)/\lambda_{\text{}},
 \end{equation}
 which has already been discussed in Ref.\ \cite{Rudge2023}. Additionally, in Fig.~\ref{fig:friction_heom_negf_noneq_friction_correlationfunction}, we observe that there is still a peak close to $x_\text{mm}$, which is now significantly more pronounced than the equilibrium case. 

 This can be explained considering the specific form of the metal-molecule coupling for this system. At $x_{\text{mm}}$, the metal-molecule couplings are equal and the electronic level sits well within the bias window, such that the electronic level is half filled. From Fig.\ \ref{fig:ml_coupling}, we see that a small positive displacement in $x$ reduces the coupling to the left lead while increasing the coupling to the right lead. Given the direction of the voltage bias, the molecule immediately decharges. Similar, a small negative displacement results in coupling only to the left lead, such that the molecule immediately charges. These two features combine for a new set of EHP creation processes, which manifest as a large $\frac{\partial \sigma^{(0),\mathcal{W}}_{\text{ss}}}{\partial x}$. 

Finally in this subsection, we note that the fluctuation-dissipation theorem is broken out of equilibrium, such that $\gamma(x) \neq D(x)/k_{B}T$. While friction describes vibrational relaxation via coupling to EHP creation, the stochastic force, and thus its correlation function, describe vibrational excitation not only via coupling to EHP relaxation but also to the electric current. For this reason, $D(x)$ displays a broad peak between $x_{\mu_\text{R}}$ and $x_{\mu_\text{L}}$, as for these coordinates the electronic level is within the bias window and there is a nonzero electric current. For most of this coordinate range, $D(x)/k_{B}T > \gamma(x)$, which is a symptom of Joule heating. Around $x_{\text{mm}}$, however, the strong peak in the friction is larger than the corresponding peak in the correlation function of the stochastic force, indicating that the vibrational relaxation processes at this point outweigh the heating originating from the electric current. 

\subsection{Interacting Vibronic Model}

Here, we consider a strongly interacting system, for which one cannot calculate the electronic forces exactly using NEGFs. In contrast, the HEOM-LD approach to electronic forces remains numerically exact even for interacting systems, demonstrating the usefulness of the extended HEOM-LD theory. Specifically, we consider a molecular model in which an electronic level is linearly coupled to one low-frequency vibrational mode, which we will treat classically, and one high-frequency vibrational mode, which we will treat quantum mechanically alongside the electronic DOFs. The molecular Hamiltonian is given by
\begin{equation}\label{ham_qu_cl}
\begin{aligned}
        H^{}_{\text{mol}}&=\left(\epsilon_{0} + \lambda_{\text{cl}} \hat x_{\text{cl}}\right)d^\dagger d +\frac{1}{2}\omega_{\text{cl}} \left(\hat x_{\text{cl}}^2+ \hat p_{\text{cl}}^2 \right) 
        \\
        &~~~+ \lambda_{\text{qu}}\hat{y}_{\text{qu}}d^\dagger d+ \frac{1}{2}\omega_{\text{qu}} \left(\hat{y}_{\text{qu}}^2+ \hat{p}_{\text{qu}}^2 \right),
\end{aligned}
\end{equation}
where $\epsilon_{0}$ is the energy of the electronic level, $\omega_{\text{cl}}$ and $\omega_{\text{qu}}$ are the frequencies of the classical and quantum vibrational modes, respectively, and  $\lambda_{\text{cl}}$ and $\lambda_{\text{qu}}$ are the corresponding electronic-vibrational couplings.

Such a model can describe a situation where a molecule is coupled not only to the electrodes but also to a high-frequency cavity mode \cite{Chen2019}, or a molecule with multiple modes operating on different timescales coupled to the transport \cite{Rudge2024}. For example, one could consider a molecular nanojunction consisting of a large molecule with a small side group attached to it. The center-of-mass motion is slow due to the large effective mass, while the frequency of the side group bond is much higher. 

To connect with the noninteracting models considered in the previous section, we again choose the metal-molecule coupling given in Eq.~(\ref{mm_coupling_andre}). Furthermore, in this subsection, we use a bandwidth of $W_\alpha=10~\text{eV}$, and set the electronic energy, $\epsilon_{0}$, such that the energy after small polaron shift is $\epsilon = \epsilon_0 - \frac{\lambda_{\text{qu}}^{2}}{\omega_{\text{qu}}} = 0.05\text{ eV}$, which is the same as in the previous section. To ensure that the timescale separation assumption is satisfied, we choose $\omega_{\text{qu}} = 300\text{ meV} = 10~\omega_{\text{cl}}$ and $\lambda_{\text{qu}} = 550\text{ meV}$. All other parameters are the same as in the previous subsection. 

\begin{center}
\begin{figure}
    \includegraphics[trim=10. .0 0 .0, clip, scale=.51]{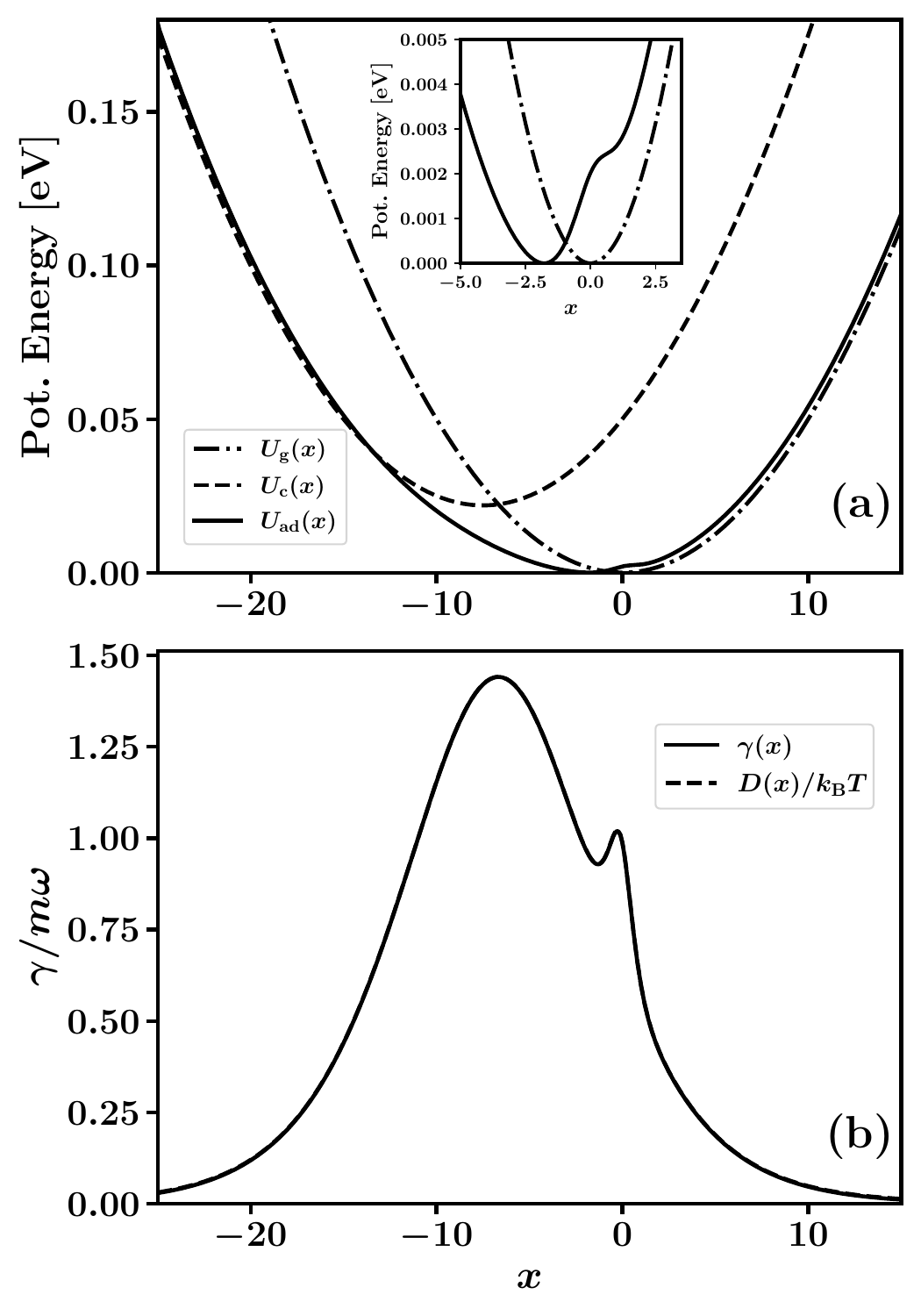}
         \centering     \phantomsubfloat{\label{fig:potentials_qu_potentials}}     \phantomsubfloat{\label{fig:potentials_qu_friction}}
    \vspace{-2\baselineskip}
    \caption{Potential energy surfaces of the classical vibrational DOFs for the interacting system (a), and friction and correlation function (b). The parameters are $\epsilon=0.05 ~\text{eV}$, $\lambda_{\text{cl}}=0.0075 ~\text{eV}$, $\omega_{\text{cl}}=0.001~\text{eV}$, $\lambda_{\text{qu}}= 0.55~\text{eV}$, and $\omega_{\text{qu}}=0.3 ~\text{eV}$. 
    }    
    \label{fig:potentials_qu}  
\end{figure}    
\end{center}

We consider first the equilibrium case, $\Phi=0\text{ V}$. Fig.~\ref{fig:potentials_qu_potentials} shows the potential energy surfaces of the neutral state, $U_\text{g}(x)$, charged state, $U_\text{c}(x)$, and the adiabatic potential, $U_\text{ad}(x)$ for this system. We observe that the addition of a quantum vibrational mode for this system barely influences the adiabatic potential.
In contrast, the nonadiabatic forces are strongly influenced by the addition of a quantum vibrational mode which can be see in Fig.~\ref{fig:potentials_qu_friction}, which
shows the friction and the correlation function for this system. 

We note that the fluctuation-dissipation theorem from Eq.~(\ref{fluctuation_dissipation_formula}) is also fulfilled for the interacting system. Again, we observe the double peak structure of the friction, which can be explained in the same way as for the noninteracting model in Subsec.~\ref{subsec:verify}. The peak at $ x_{\varepsilon_{\text{F}}}=-(\epsilon - \varepsilon_{\text{F}})/\lambda_{\text{cl}}$ is at the position where the electronic level crosses the chemical potential. In contrast, the second peak originates from the position dependency of the metal-molecule coupling. Here, at $ x_{{\text{c}}}$, the spatial derivatives of the metal-molecule couplings reach their extrema, resulting in the peak in the friction at this position. 

We note that the electronic friction of the interacting system in Fig.~\ref{fig:potentials_qu_friction} is significantly larger than the electronic friction of the noninteracting system for the entire range of vibrational coordinates (cf. Fig.~\ref{fig:potentials_cl_friction}). Even at equilibrium, the exchange of quanta with the high-frequency mode allows additional lead electrons to transfer to the molecular junction, absorb vibrational energy from the classical mode, and transfer back to a higher energetic state in the leads. Thus, the addition of a quantum vibrational mode greatly enhances the number of EHP creation process, resulting in a larger electronic friction for the interacting system compared to the noninteracting system \cite{Rudge2023,Hrtle2011}. 

However, while for the interacting system the peak at $ x_{{\text{c}}}$ in Fig.~\ref{fig:potentials_qu_friction} has a smaller value than the peak at $ x_{\mu_{\text{}}}$, for the corresponding peaks in the noninteracting system in Fig.~\ref{fig:potentials_cl_friction} the opposite is true and can be explained as in the following. 
The enhancement of the electron-hole pair creation process is especially dominant in the vicinity of $ x_{\mu}$ where the electronic level crosses the chemical potential. Therefore, it influences the friction at $ x_{\mu}$ more than at $x_\text{mm}$. Hence, for the interacting system in Fig.~\ref{fig:potentials_qu_friction}, the peak in the friction at around $ x_\text{mm}$ is less pronounced compared to the peak at around $ x_{\mu}$, in contrast to noninteracting case in Fig.~\ref{fig:potentials_cl_friction}. 

\begin{center}
\begin{figure}[t]
    \includegraphics[trim=15. .0 9 .0, clip, scale=.54]{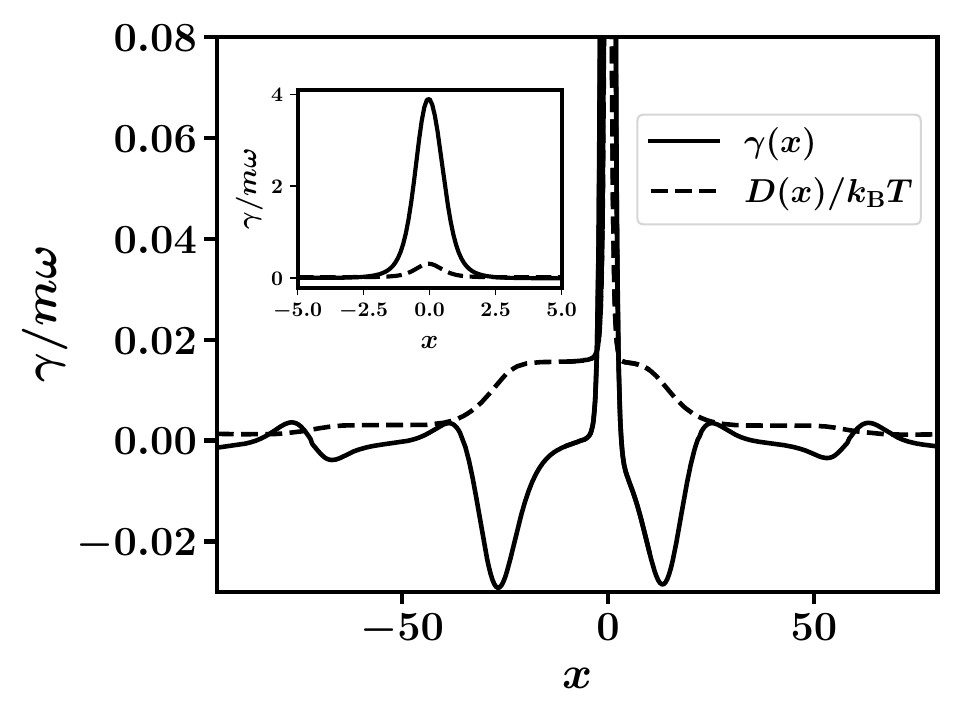}
    \caption{Friction and the correlation function for the interacting system out of equilibrium. The bias voltage is $\Phi=1 ~\text{V}$. All other parameters are the same as in Fig.~\ref{fig:potentials_qu}. }
    \label{fig:friction_qu_cl_1v}
\end{figure}    
\end{center}

We now consider the nonequilibrium case and apply a bias voltage $\Phi=1~\text{V}$ to the leads. Figure ~\ref{fig:friction_qu_cl_1v} shows the friction and the correlation function for the interacting system. Out of equilibrium, the peak in the friction in Fig.~\ref{fig:potentials_qu_friction} at $x_{\mu_{\text{}}}$ splits into two peaks at $ x_{\mu_{\text{L}}}$ and $x_{\mu_{\text{R}}}$. Again we observe that the peak in the friction close to $x_\text{mm}$ is significantly more pronounced in comparison to the other peaks than in the equilibrium situation. This can be explained in the same way as in the previous subsection considering the specific form of the metal-molecule coupling for this system. 

Figure~\ref{fig:friction_qu_cl_1v} shows additional features, which can only be observed for the interacting system. These additional extrema in the friction appear at positions \cite{Rudge2023}
\begin{equation}
   x_{\mu_\alpha,n}= \frac{\mu_{\alpha} - \left(\epsilon_0-\varepsilon_\text{F} - {\lambda_{\text{qu}}^2 }/{\omega_{\text{qu}}} + n\omega_{\text{qu}}\right)} { \lambda_{\text{cl}}}. 
\end{equation}
At these positions, EHP creation and annihilation processes mediated by the exchange of $n$ vibrational quanta with the high-frequency mode are resonant, resulting in additional extrema of the friction compared to the noninteracting case. Most interestingly, for this interacting system, such processes do not always result in a cooling of the low-frequency mode, as the electronic friction is negative for a significant portion of the coordinate range, which indicates areas of space in which local heating effects dominate. The appearance of such negative friction could connect to vibrational instabilities within molecular nanojunctions \cite{L2010, Todorov2010, L2011}. 

\section{Conclusion} \label{sec: Conclusion}

In this work, we extended the theory of electronic friction within the HEOM formalism to systems with a position-dependent metal-molecule coupling. The incorporation of such couplings is critical for a realistic description of many scenarios, such as reaction dynamics at metal surfaces, quantum shuttling, and vibrational instabilities in molecular nanojunctions. Consequently, the updated theory represents one of the most general mixed quantum-classical approaches to the nonadiabatic dynamics of molecules interacting with metal surfaces currently available.

Generally, our theory showed that incorporating a position-dependent metal-molecule coupling adds extra terms to the adiabatic contribution to the mean force, the electronic friction, and the correlation function of the stochastic force. We next demonstrated that such extra terms can critically affect both the adiabatic and nonadiabatic forces. We first applied the updated HEOM-LD approach to a nonequilibrium charge transport through a simple vibronic system containing a harmonic vibrational mode linearly coupled to an electronic level. Since the electronic part of this system is noninteracting, we were also able to compare our updated theory with electronic friction obtained from NEGFs, verifying our extended approach. We next demonstrated the power of the HEOM-LD approach by calculating the electronic forces of an interacting two-mode system, where one of the modes was treated classically and the other quantum mechanically. 

In both scenarios, we demonstrated that incorporating a position-dependent metal-molecule coupling significantly affects the electronic forces at positions where the spatial derivatives of the metal-molecule couplings reach their extrema. This was particularly relevant for the electronic friction and stochastic force, which are critical for understanding nonadiabatic processes at metal surfaces. Beyond the reduced models considered here, we foresee great applicability of the updated HEOM-LD approach to more realistic models, as the HEOM treats the quantum mechanical part of the problem in a numerically exact manner, while the Langevin dynamics is capable of handling anharmonic vibrational degrees of freedom with nonlinear electronic-vibrational interactions. 

\section*{Acknowledgment}
This work was supported by the German Research Foundation (DFG) through Research Unit FOR5099. S.L.R. thanks
the Alexander von Humboldt Foundation for support via a research fellowship. Moreover, the authors thank Riley Preston for helpful discussions.  Furthermore, support by the state of
Baden-Württemberg through bwHPC and the DFG through Grant No. INST 40/575-1 FUGG (JUSTUS 2 cluster) is gratefully acknowledged.

\appendix

\section{Definition of Wigner-transformed Superoperators in the HEOM Formalism}\label{appendix:a}

The superoperators $\mathcal{\tilde{A}}^{\mathcal{W}}$, $ \mathcal{\tilde{A}}_{\partial x_i}^{\mathcal{W}}$, and $\mathcal{\tilde{\kappa}}$ introduced in Eq.~(\ref{drhodt_new}) are defined by their action on the vector of all ADOs, 
 \begin{equation}
    \mathcal{\tilde{A}}^{\mathcal{W}}\begin{bmatrix}
        \rho^{(0),\mathcal{W}} \\
        \rho_{j_0}^{(1),\mathcal{W}} \\
        \vdots \\
        \rho_{\bold j}^{(n),\mathcal{W}}
        \\
        \vdots
    \end{bmatrix}=\begin{bmatrix}
         -i \sum_j \mathcal{{A}}^{\Bar{\sigma},\mathcal{W}}_{\alpha,m} \rho_{j}^{(1),\mathcal{W}}
         \\
        - i \sum_j \mathcal{{A}}^{\Bar{\sigma},\mathcal{W}}_{\alpha,m} \rho_{j j_0}^{(2),\mathcal{W}}\\
        \vdots\\
        
       -i\sum^{}_{j}         \mathcal{{A}}^{\Bar{\sigma},\mathcal{W}}_{\alpha,m} \rho_{\bold{j}^+}^{(n+1),\mathcal{W}}
        \\
        \vdots
    \end{bmatrix},
\end{equation}
 \begin{equation}
    \mathcal{\tilde{A}}_{\partial x_i}^{\mathcal{W}}\begin{bmatrix}
        \rho^{(0),\mathcal{W}} \\
        \rho_{j_0}^{(1),\mathcal{W}} \\
        \vdots \\
        \rho_{\bold j}^{(n),\mathcal{W}}
        \\
        \vdots
    \end{bmatrix}=\begin{bmatrix}
         -i \sum_j \left(\mathcal{{A}}^{\Bar{\sigma},\mathcal{W}}_{\alpha,m}\right)'_i\rho_{j}^{(1),\mathcal{W}}
         \\
        - i \sum_j \left(\mathcal{{A}}^{\Bar{\sigma},\mathcal{W}}_{\alpha,m}\right)'_i \rho_{j j_0}^{(2),\mathcal{W}}\\
        \vdots
        \\
       -i\sum^{}_{j}         \left(\mathcal{{A}}^{\Bar{\sigma},\mathcal{W}}_{\alpha,m}\right)'_i \rho_{\bold{j}^+}^{(n+1),\mathcal{W}}
        \\\vdots
    \end{bmatrix},
\end{equation}

and
\begin{equation}
    \mathcal{\tilde{\kappa}}\begin{bmatrix}
        \rho^{(0),\mathcal{W}} \\
        \rho_{j_1}^{(1),\mathcal{W}} \\
        \vdots \\
        \rho_{\bold j^{}}^{(n),\mathcal{W}}
        \\
        \vdots
    \end{bmatrix}=\begin{bmatrix}
        0\\
         \mathcal{{\kappa}}_{j_1}\rho_{j_1}^{(1),\mathcal{W}}\\
        \vdots\\
            \sum_{r=1}^{n} \kappa_{j_r}\rho_{\bold j^{}}^{(n),\mathcal{W}}
            \\
            \vdots
    \end{bmatrix},
\end{equation}
where we have used the notation introduced in Eq.~(\ref{notation_1}) and Eq.~(\ref{notation_2}). 

\section{Time Evolution of the Phase-Space Probability Density in The HEOM Formalism}
\label{appendix:b}

 In the following, we show a detailed derivation of the time evolution of the PSPD from Eq.~(\ref{PSPD_final_vec}) in terms of the HEOM. We follow a similar procedure as in Ref.~\cite{Jin2008}, and compare the time evolution of the molecular density matrix with the time evolution of the zeroth tier ADO. In contrast to Ref.~\cite{Jin2008}, however, we consider the time evolution of the Wigner-transformed operators. 
 
 We start by considering the time evolution of the molecular density matrix, given by the Wigner-transformed Liouville-von Neumann equation
\begin{equation}\label{wigner_transformed_liouville_eq}
    \begin{aligned}
        \frac{\partial}{\partial t} \rho_{\text{mol}}^{\mathcal{W}}
                 = \:\:\: \text{tr}_{\text{met}}\Bigg( & -i\Bigg. \left[H^{\mathcal{W}}_{\text{mol}},\rho^{\mathcal{W}}_{\text{T}} \right] -i\left[H^{\mathcal{W}}_{\text{met-mol}},\rho^{\mathcal{W}}_{\text{T}} \right] + \\
                 & \big \{H^{\mathcal{W}}_{\text{mol}},\rho^{\mathcal{W}}_{\text{T}} \big\}_{\text{a}} +\big \{H^{\mathcal{W}}_{\text{met-mol}},\rho^{\mathcal{W}}_{\text{T}} \big\}_{\text{a}}
         \Bigg). 
    \end{aligned}
\end{equation}
This is equivalent to the time evolution of the Wigner-transformed $0$th-tier ADO:
\begin{equation} \label{dt_0th_ado}
    \begin{aligned}
        \frac{\partial}{\partial t} \rho^{(0),\mathcal{W}}_{}
        = \:\:\: & -i\left[H^{\mathcal{W}}_{\text{mol}},\rho^{(0),\mathcal{W}}_{\text{}} \right]  +\big\{H^{\mathcal{W}}_{\text{mol}},\rho^{(0),\mathcal{W}}_{\text{}} \big\}_{\text{a}} \\
        & -i  \left( \sum_j \mathcal{A}_{m,\alpha}^{\bar \sigma} \rho_{j}^{(1)} \right)^\mathcal{W},
    \end{aligned}
\end{equation}
where the Wigner transform of the coupling-up superoperator is evaluated explicitly as
\begin{widetext}
  \begin{equation}\label{wigner_A}
    \begin{aligned}
        \left( \sum_j \mathcal{A}_{m,\alpha}^{\bar \sigma} \rho_{j}^{(1)} \right)^\mathcal{W}
        = & \sum_j t_{\alpha,m}\left( g^\ph_{\alpha}(\hat{\mathbf{x}}) d_m^{\bar \sigma}\rho_{j}^{(1)} - \right. 
\left. \rho_{j}^{(1)} g^\ph_{\alpha}(\hat{\mathbf{x}}) d_m^{\bar \sigma}\right)^\mathcal{W} \\
 = & \sum_j    g_{\alpha}(\mathbf{x})t_{\alpha,m}  \left(d_m^{\bar \sigma}\rho_{j}^{(1),\mathcal{W}} - \rho_{j}^{(1),\mathcal{W}} d_m^{\bar \sigma}\right) -
\frac{i}{2}\sum_j \sum^N_{i=1} \frac{\partial g_{\alpha}}{\partial x_i} t_{\alpha,m}
 \Big(d_m^{\bar \sigma}\frac{\partial \rho_{j}^{(1),\mathcal{W}}}{\partial p_i} -  \frac{\partial \rho_{j}^{(1),\mathcal{W}}}{\partial p_i}   d_m^{\bar \sigma}\Big) .
    \end{aligned}
\end{equation}  
Now, we relate the different terms in Eq.~(\ref{wigner_transformed_liouville_eq}) to their respective counterparts in Eq.~(\ref{dt_0th_ado})

 \begin{equation}\label{wigner_A_2}
    \begin{aligned}
        -i \left( \sum_j \mathcal{A}_{m,\alpha}^{\bar \sigma} \rho_{j}^{(1)} \right)^\mathcal{W}
        = \:\:\: \text{tr}_{\text{met}}\Bigg( & i\left[H^{\mathcal{W}}_{\text{met-mol}},\rho^{\mathcal{W}}_{\text{T}} \right] +
         \big \{H^{\mathcal{W}}_{\text{met-mol}},\rho^{\mathcal{W}}_{\text{T}} \big\}_{\text{a}}
         \Bigg).
        \end{aligned}
\end{equation}   
\end{widetext}
We write out the $\{ \dots \}_\text{a}$ brackets on the right hand side of Eq.~(\ref{wigner_A_2}), and match the terms on both sides of Eq.~(\ref{wigner_A_2}) depending on 
$ \frac{\partial }{\partial p_i}\rho_{j}^{(1),\mathcal{W}}$ and $\frac{\partial}{\partial p_i} \rho^{\mathcal{W}}_{\text{T}}$, 
    \begin{equation}\label{matching_1}
    \begin{aligned} 
    & \sum_j \sum^N_{i=1}  \frac{\partial g_{\alpha}}{\partial x_i} t_{\alpha,m} \left( d_m^{\bar \sigma}\frac{\partial \rho_{j}^{(1),\mathcal{W}}}{\partial p_i} - \frac{\partial \rho_{j}^{(1),\mathcal{W}}}{\partial p_i}  d_m^{\bar \sigma}\right)
  = \\
   & \:\:\:\:\: \text{tr}_{\text{met}}\left[ \sum^N_{i=1}\left(\frac{\partial H^{\mathcal{W}}_{\text{met-mol}}}{ \partial x_i}  \frac{\partial \rho^{\mathcal{W}}_{\text{T}} }{ \partial p_i} - \frac{\partial \rho^{\mathcal{W}}_{\text{T}}}{  \partial p_i}  \frac{\partial  H^{\mathcal{W}}_{\text{met-mol}}}{ \partial x_i}  \right)\right].
    \end{aligned}
\end{equation}
Writing out the explicit form of $H^{\mathcal{W}}_{\text{met-mol}}$ and comparing both sides in Eq.~(\ref{matching_1}) then implies
\begin{equation}\label{trace_bath_dagger_rho}
    \begin{aligned}
    \sum_{j}    \frac{\partial g_{\alpha}}{\partial x_i}t_{\alpha,m} 
 d_m^{\bar \sigma}\frac{\partial \rho_{j}^{(1),\mathcal{W}} }{\partial p_i}
   &= 
   \sum_{\alpha, m} \frac{\partial g_{ \alpha}}{\partial x_i}t_{\alpha,m}\text{tr}_{\text{met}}\left( c^{\sigma}_{ \alpha}d_m^{\bar \sigma}\frac{\partial \rho^{\mathcal{W}}_{\text{T}}}{\partial p_i}\right).
    \end{aligned}
\end{equation}
Equation~(\ref{trace_bath_dagger_rho}) allows us to express the trace over the metal DOFs in terms of the $1$st-tier ADOs. 

We now consider again the expression for the time evolution of the PSPD from Eq.~(\ref{dAdt_comm}). The first term on the right hand side of Eq.~(\ref{dAdt_comm}) can be written as
\begin{equation} \label{trace_SB}
\begin{aligned}
     \text{tr}_{\text{mol}^{\text{qu}}+\text{met}} \left[ \big \{H^{\mathcal{W}}_{\text{mol}},\rho_{\text{T}} \big\}_\text{a}\right]     
     &=
     \text{tr}_{\text{mol}^{\text{qu}}}  \left[\big\{ H^{\mathcal{W}}_{\text{mol}}, \rho^{{\mathcal{W}}}_{\text {mol}}\big \}_\text{a} \right].
\end{aligned}   
\end{equation}
By exploting the connection between the trace over the metal DOFs and the first tier ADOs from Eq.~(\ref{trace_bath_dagger_rho}), we can write the second term on the right hand side of Eq.~(\ref{dAdt_comm}) as
\begin{widetext}
    \begin{equation}\label{trace_bath_c_rho_ADO}
    \begin{aligned} 
    \text{tr}_{\text{mol}^{\text{qu}}+\text{met}} \left[  \{H^{\mathcal{W}}_{\text{met-mol}},\rho^{{\mathcal{W}}}_T \}_\text{a}\right]        
        &=\text{tr}_{\text{mol}^{\text{qu}}}  \Bigg[ \Bigg.  \sum_{\alpha,  m}\sum^N_{i=1}\frac{\partial g_{\alpha}(\mathbf{x})}{\partial x_i}   t_{\alpha,m} 
    \left(  d_m^{\dagger}~\text{tr}_{\text{met}} \left[c_{\alpha} \frac{\partial \rho^{{\mathcal{W}}}_\text{T}}{\partial p_i}\right]   +  d_m~ \text{tr}_{\text{met}} \left[  c_{\alpha}^{\dagger} \frac{\partial \rho^{{\mathcal{W}}}_\text{T}}{\partial p_i}\right]  \right)\Bigg] \Bigg. 
        \\
    &= \text{tr}_{\text{mol}^ \mathcal{W}} \left[\sum_{\alpha,\sigma,l,m}  \sum^N_{i=1} \frac{\partial g_{\alpha}(\mathbf{x})}{\partial x_i} t_{\alpha,m}  d_m^{\bar \sigma}\frac{\partial}{\partial p_i}\rho_{j}^{(1),\mathcal{W}} \right].
        \\
    \end{aligned}
\end{equation}
\end{widetext}
Finally, adding Eq.~(\ref{trace_SB}) and Eq.~(\ref{trace_bath_c_rho_ADO}), the time evolution of the PSPD can be written as
\begin{equation}\label{PSPD_final_appendix}
\begin{aligned}
                    \frac{\partial A}{\partial t}
                    = \:\:\: &
                    \text{tr}_{\text{mol}^{\text{qu}}}  \left[\big\{ H^{\mathcal{W}}_{\text{mol}}, \rho^{\mathcal{W}}_S \big \}_\text{a}\right]
                    + \\
                    & \text{tr}_{\text{mol}^{\text{qu}}}  \left[\sum_{j} \sum^N_{i=1}  \frac{\partial g_{\alpha} }{\partial x_i} t_{\alpha,m} d_m^{\bar \sigma}\frac{\partial \rho_{j}^{(1),\mathcal{W}}}{\partial p_i} \right].
\end{aligned}
\end{equation}
Considering the vector off all ADOs introduced in Eq.~(\ref{ADO_vec}), and the superoperator 
${\mathcal{H}}_{\text{met-mol},i}^{\mathcal{W}}$, which has been introduced in Eq.~(\ref{h_sb_definition}), we obtain the expression in Eq.~(\ref{PSPD_final_vec}).
 
\section{Derivation of the Fokker-Planck Equation }
\label{appendix:c}
In the following, we show that by assuming a timescale separation between the dynamics of classical and quantum DOFs the time evolution of the PSPD in Eq.~(\ref{PSPD_final_vec}) is given by a Fokker-Planck equation.

We assume that the dynamics of the quantum mechanical DOFs and the classical DOFs take place on timescales $\tau_\text{qu}$ and $\tau_\text{cl}$, respectively. We make the same assumption as in Ref.~\cite{Rudge2023}, that the dynamics of the classical DOFs are considerably slower than the quantum dynamics, so that $\tau_\text{cl}   \gg \tau_\text{qu}$. Hence, on a timescale determined by $\tau_\text{cl}$, the quantum DOFs will immediately reach their steady state. Moreover, within the timescale determined by $\tau_\text{qu}$, the classical vibrational DOFs will barely influence the quantum DOFs. Therefore, we can write the ADOs as in Eq.~(\ref{rho_ss}).
Recalling the time evolution of the PSPD in terms of the HEOM from Eq.~(\ref{PSPD_final_vec}) and using Eq.~(\ref{rho_ss}), we obtain
   \begin{equation}\label{dAdt_B}
    \begin{aligned}
                    \frac{\partial A}{\partial t} = \: & \text{tr}_{\text{mol}^{\text{qu}}}  \left[ \big\{ \big\{{H}^{\mathcal{W}}_{\text{mol}}, \text{\boldmath$\rho$}^{\mathcal{W}} \big\}\big\}_\text{a}\right]+ \\
                    & \text{tr}_{\text{mol}^{\text{qu}}}  \left[ \sum^N_{i=1}{\mathcal{H}}_{\text{met-mol},i}^{\mathcal{W}} \frac{\partial \text{\boldmath$\rho$}^{\mathcal{W}} }{\partial p_i} \right],
    \end{aligned}
\end{equation} 
which, when we insert the ansatz from Eq.~(\ref{rho_ss}), becomes 
\begin{equation}\label{dAdt_B}
    \begin{aligned}
                    \frac{\partial A}{\partial t}  = & \text{tr}_{\text{mol}^{\text{qu}}}  \left[ \big\{\big\{{H}^{\mathcal{W}}_{\text{mol}}, A\text{\boldmath$\sigma$}_{\text{ss}}^{\mathcal{W}}\big \}\big\}_\text{a}\right]+\text{tr}_{\text{mol}^{\text{qu}}}  \left[ \big\{\big\{{H}^{\mathcal{W}}_{\text{mol}}, \bold B\big \}\big\}_\text{a}\right] + \\
                    & \text{tr}_{\text{mol}^{\text{qu}}}  \left[ \sum^N_{i=1}{\mathcal{H}}_{\text{met-mol},i}^{\mathcal{W}} \frac{\partial}{\partial p_i}A\text{\boldmath$\sigma$}_{\text{ss}}^{\mathcal{W}}\right]+\\
                    & \text{tr}_{\text{mol}^{\text{qu}}}  \left[ \sum^N_{i=1}{\mathcal{H}}_{\text{met-mol},i}^{\mathcal{W}} \frac{\partial}{\partial p_i} \bold B\right].
    \end{aligned}
\end{equation} 
We now write out the antisymmetrized Poisson brackets explicitly:
\begin{equation}\label{dAdt_B}
    \begin{aligned}
                    \frac{\partial A}{\partial t} = &  -\sum^N_{i=1} \Big( \frac{p_i}{m_i} \frac{\partial A}{ \partial x_i} + 
                    \text{tr}_{\text{mol}^{\text{qu}}}  \big[ \frac{\partial H^{\mathcal{W}}_{\text{mol}}}{\partial x_i} \text{\boldmath$\sigma$}^{\mathcal{W}}_{\text{ss}}\big] \frac{\partial A}{\partial p_i} + \\
        & \text{tr}_{\text{mol}^{\text{qu}}}  \left[ \frac{\partial H^{\mathcal{W}}_{\text{mol}}}{\partial x_i}  \frac{\partial  \bold B}{\partial p_i}\right] + \text{tr}_{\text{mol}^{\text{qu}}}  \left[ {\mathcal{H}}_{\text{met-mol},i}^{\mathcal{W}} \frac{\partial}{\partial p_i}A\text{\boldmath$\sigma$}_{\text{ss}}^{\mathcal{W}}\right] + \\
        &         \text{tr}_{\text{mol}^{\text{qu}}}  \left[ {\mathcal{H}}_{\text{met-mol},i}^{\mathcal{W}} \frac{\partial}{\partial p_i} \bold B\right] \Big],
    \end{aligned}
\end{equation} 

Next, we rearrange Eq.~(\ref{rho_ss}) to obtain the time evolution of $\bold B $,

\begin{equation}\label{dBdt_new}
\begin{aligned}
        \frac{\partial}{\partial t} \bold B&= \frac{\partial}{\partial t}\text{\boldmath$\rho$}^{\mathcal{W}} -\frac{\partial}{\partial t}  A \text{\boldmath$\sigma$}^{\mathcal{W}}_{\text{ss}}.
\end{aligned}
\end{equation}

which we can express in terms of the Wigner-transformed HEOM from Eq.~(\ref{drhodt_new}):
\begin{equation}\label{dBdt_via_heom}
    \begin{aligned}    
    \frac{\partial}{\partial t} \bold B
        &=
        -\mathcal{L}^{\mathcal{W}}\bold B
          +\big \{ H^{\mathcal{W}}_{\text{mol}},\left( A\text{\boldmath$\sigma$}^{\mathcal{W}}_{\text{ss}}+\bold B \right)\big \}_\text{a} 
          \\
          &~~~~- \frac{i}{2}   \sum^N_{i=1} \left(\mathcal{\Tilde{{C}}}_{\partial {x}_i}^{\mathcal{W}} +\mathcal{\Tilde{{A}}}_{\partial x_i}^{\mathcal{W}}\right)\frac{\partial}{\partial p_i}\left( A\text{\boldmath$\sigma$}^{\mathcal{W}}_{\text{ss}}+\bold B \right) -\frac{\partial}{\partial t}  A \text{\boldmath$\sigma$}^{\mathcal{W}}_{\text{ss}},
    \end{aligned}
\end{equation}
where we have also used Eq.~(\ref{Liouvillian}). We now formally solve Eq.~(\ref{dBdt_via_heom}) for $\bold B$, obtaining 

\begin{equation}\label{B_via_heom}
    \begin{aligned}    
   \bold B = \: & (\mathcal{L}^{\mathcal{W}})^{-1}\Bigg(-\frac{\partial }{\partial t}\bold B  
          +\big \{ H^{\mathcal{W}}_{\text{mol}},\left( A\text{\boldmath$\sigma$}^{\mathcal{W}}_{\text{ss}}+\bold B \right)\big \}_\text{a} - \\
& \frac{i}{2}   \sum^N_{i=1} \left(\mathcal{\Tilde{{C}}}_{\partial {x}_i}^{\mathcal{W}} +\mathcal{\Tilde{{A}}}_{\partial x_i}^{\mathcal{W}}\right)\frac{\partial}{\partial p_i}\left( A\text{\boldmath$\sigma$}^{\mathcal{W}}_{\text{ss}}+\bold B\right)-\frac{\partial}{\partial t}  A \text{\boldmath$\sigma$}^{\mathcal{W}}_{\text{ss}}\Bigg).
    \end{aligned}
\end{equation}
We note that the pseudoinverse of the joint Liouville operator, $\left(\mathcal{L}^{\mathcal{W}}(\mathbf{x})\right)^{-1}$, can be calculated via Laplace transform 
\begin{equation}
    \left(\mathcal{L}^{\mathcal{W}}(\mathbf{x})\right)^{-1}=\lim_{\delta \to 0_+} \int_0^\infty dt ~e^{-\left(\mathcal{L}^{\mathcal{W}}(\mathbf{x})+\delta\right)t}.
\end{equation}
Now we express $\frac{\partial}{\partial t} A $ in Eq.~(\ref{B_via_heom}) through Eq.~(\ref{dAdt_B}), and insert the whole expression for $\bold B$ into Eq.~(\ref{dAdt_B}). Thereby, we assume $\frac{\partial}{\partial t} \bold B \ll\frac{\partial}{\partial t}  A \text{\boldmath$\sigma$}^{\mathcal{W}}_{\text{ss}}$, $\frac{\partial}{\partial x_i} \bold B \ll \frac{\partial}{\partial x_i} \left(A \text{\boldmath$\sigma$}^{\mathcal{W}}_{\text{ss}} \right)$, and
$\frac{\partial}{\partial p_i} \bold B \ll \frac{\partial}{\partial p_i} A \text{\boldmath$\sigma$}^{\mathcal{W}}_{\text{ss}}$, and that, therefore, derivatives of $\bold B$ can be neglected. Under this assumption, it becomes apparent that only four terms contribute to $\bold B$, so that 
\begin{equation}\label{B_contribution}
    \begin{aligned}
        \bold B = & \:\:-\sum^N_{i=1} \frac{p_i }{m_i} \left(\mathcal{L}^{\mathcal{W}}\right)^{-1} \frac{\partial \text{\boldmath$\sigma$}_{\text{ss}}^{\mathcal{W}}}{\partial x_i}A +
        \\
        & \frac{1}{2}\left(\mathcal{L}^{\mathcal{W}}\right)^{-1} \sum^N_{i=1} \left( \frac{\partial H^{\mathcal{W}}_{\text{mol}}}{\partial x_i}\text{\boldmath$\sigma$}_{\text{ss}}^{\mathcal{W}}-\text{\boldmath$\sigma$}_{\text{ss}}^{\mathcal{W}} \frac{\partial H^{\mathcal{W}}_{\text{mol}}}{\partial x_i}\right)\frac{\partial A}{\partial p_i} -
        \\
        & \frac{i}{2} \left(\mathcal{L}^{\mathcal{W}}\right)^{-1}\sum^N_{i=1} \left(\mathcal{\Tilde{{C}}}_{\partial {x}_i}^{\mathcal{W}} +\mathcal{\Tilde{{A}}}_{\partial x_i}^{\mathcal{W}}\right)\text{\boldmath$\sigma$}_{\text{ss}}^{\mathcal{W}}\frac{\partial A}{\partial p_i} +
        \\
        &\left(\mathcal{L}^{\mathcal{W}}\right)^{-1} \sum^N_{i=1}\Bigg(-\frac{p_i}{m_i} \frac{\partial A}{ \partial x_i} + \text{tr}_{\text{mol}^{\text{qu}}}  \left[ \frac{\partial H^{\mathcal{W}}_{\text{mol}}}{\partial x_i} \text{\boldmath$\sigma$}^{\mathcal{W}}_{\text{ss}}\right] \frac{\partial A}{\partial p_i}
+\\
& \hspace{2.5cm}\text{tr}_{\text{mol}^{\text{qu}}}  \left[ {\mathcal{H}}_{\text{met-mol},i}^{\mathcal{W}} \frac{\partial}{\partial p_i}A\text{\boldmath$\sigma$}_{\text{ss}}^{\mathcal{W}}\right] \Bigg)\text{\boldmath$\sigma$}_{\text{ss}}^{\mathcal{W}}.
\end{aligned}
\end{equation}
The first two terms on the right hand side of Eq.~(\ref{B_contribution}) come from the antisymmetrized Poisson bracket in Eq.~(\ref{B_via_heom}) and also appear in the constant metal-molecule coupling case. The third term comes from the inclusion of a $\mathbf{x}$-dependency in the metal-molecule coupling, and the last term comes from inserting Eq.~(\ref{dAdt_B}) into Eq.~(\ref{B_via_heom}) and neglecting derivatives of $\bold B$. Inserting these contributions in the last line of Eq.~(\ref{dAdt_B}), we obtain the Fokker-Planck equation from Eq.~(\ref{fokker_planck_equation}). 

\bibliography{main.bib}
\end{document}